\documentclass[12pt]{iopart}
\usepackage{latexsym}
\usepackage{graphicx}

\def\op#1{\hat{#1}}
\def\ket#1{| #1 \rangle}
\def\bra#1{\langle #1 |}
\def\ave#1{\langle #1 \rangle}
\def\vec#1{{\bf #1}}

\def\Tr{{\rm Tr}}

\def\Rb87{${}^{87}$Rb}

\newif\ifpdflatex\pdflatextrue
\makeatletter\@ifundefined{pdfoutput}{\pdflatexfalse}\makeatother
\def\myincludegraphics[#1]#2#3{%
\ifpdflatex \includegraphics[#1]{#2}
\else       \includegraphics[#1]{#3}
\fi}
\begin{document}
\bibliographystyle{prsty}
\title{Constructive control of quantum systems using factorization of unitary operators}
\author{S.~G.\ Schirmer}
\address{Quantum Processes Group and Department of Applied Maths,
         The Open University, Milton Keynes, MK7 6AA, 
         United Kingdom}
\author{Andrew D.\ Greentree}
\address{Quantum Processes Group and Department of Physics and Astronomy,
         The Open University, Milton Keynes, MK7 6AA, 
         United Kingdom}
\author{Viswanath Ramakrishna}
\address{Center for Signals, Systems and Telecommunications and 
         Dept of Mathematical Sciences , EC 35, 
         University of Texas at Dallas, Richardson, Texas 75083, USA}
\author{Herschel Rabitz}
\address{Department of Chemistry, Frick Laboratories, 
         Princeton University, Princeton, NJ 08544, USA}
\eads{\mailto{s.g.schirmer@open.ac.uk}, 
      \mailto{a.d.greentree@open.ac.uk},
      \mailto{vish@utdallas.edu}, 
      \mailto{hrabitz@princeton.edu}} 
\date{January 21, 2001} 
\begin{abstract}
We demonstrate how structured decompositions of unitary operators can be employed to 
derive control schemes for finite-level quantum systems that require only sequences 
of simple control pulses such as square wave pulses with finite rise and decay times
or Gaussian wavepackets.  To illustrate the technique, it is applied to find control 
schemes to achieve population transfers for pure-state systems, complete inversions
of the ensemble populations for mixed-state systems, create arbitrary superposition
states and optimize the ensemble average of dynamic observables.
\end{abstract}
\maketitle

\section{Introduction}
\label{sec:intro}

The ability to control quantum-mechanical systems is an essential prerequisite for many 
novel applications that require the manipulation of atomic and molecular quantum states 
\cite{SCI288p0824}.  Among the important applications of current interest are quantum 
state engineering \cite{PRA63n023408}, control of chemical reactions \cite{JCP113p03510, 
SCI292p709, SCI282p919, SCI279p1875, SCI279p1879}, control of molecular motion
\cite{PRA61n033816}, selective vibrational excitation of molecules \cite{PCCP2p1117}, 
control of rotational coherence in linear molecules \cite{PRA61n033816}, photo-dissociation
\cite{JPCA104p4882}, laser cooling of internal molecular degrees of freedom \cite{FD113p365,
PRA63n013407}, and quantum computation \cite{qph0104030, qph0103118, CPL343p633,SCI287p463,
PRA65n042301}.

Due to the wide range of applications, the immediate aims of quantum control may vary.  
However, the control objective can usually be classified as one of the following: 
\begin{enumerate}
\item \label{a}
      To steer the system from its initial state to a target state with desired properties,
\item \label{b}
      To maximize the expectation value or ensemble average of a selected observable,
\item \label{c}
      To achieve a certain evolution of the system.  
\end{enumerate}
Despite the apparent dissimilarity, these control objectives are closely related.  Indeed,
(\ref{a}) is a special case of (\ref{b}) in which the observable is the projector onto the
subspace spanned by the target state.  (\ref{b}) is a special case of (\ref{c}), where we 
attempt to find an evolution operator that maximizes the expectation value of the selected 
observable either at a specific target time or at some time in the future.  Hence, one of 
the central problems of quantum control is to achieve a desired evolution of the system by
applying external control fields, and the primary challenge is to find control pulses (or 
sequences of such pulses) that are feasible from a practical point of view and effectively 
achieve the control objective.

Many control strategies for quantum systems have been proposed.  Selective excitation 
of energy eigenstates, for instance, can be achieved using light-induced potentials and 
adiabatic passage techniques \cite{ARPC52p763, EPJD14p147, PRL85p4241, JCP114p8820}, which
have the advantage of being relatively insensitive to perturbations of the control fields
and Doppler shifts arising from atomic or molecular motion \cite{PRA61n043413,PRA63n043415}.
Efficient numerical algorithms based on optimal control techniques have been developed to
address problems such as optimization of observables for pure-state \cite{JCP109p385, 
JCP112p05081,JCP110p7142} and mixed-state quantum systems \cite{PRA61n012101,JCP110p9825}. 
Quantum feedback control using weak measurements or continuous state estimation has been
applied to quantum state control problems \cite{PRA62n022108, PRA62n012307, PRA62n012105, 
PRL85p3045, PRA60p2700, PRA49p2133}.  Learning control based on genetic or evolutionary 
algorithms \cite{JCP113p10841, JCP110p34, PRE56p3854, CP217p389, JPC99p5206, PRL68p1500} 
has been a useful tool for quantum control, especially for complex problems for which 
accurate models are not available and in experimental settings \cite{NAT406p164, APB65p779}.
Other approaches based on local control techniques \cite{JCP109p9318} or a hydrodynamical
formulation \cite{JPA33p4643} have been suggested as well, and this list is not exhaustive.

In this paper we pursue an alternative, constructive approach to address the problem of 
control of non-dissipative quantum systems.  Note that although real atomic or molecular 
systems are subject to dissipative processes due to the finite lifetimes of the excited 
states, etc., we can treat these systems as non-dissipative if we ensure that the time 
needed to complete the control process is \emph{significantly} less than the relaxation 
times.  The technique we develop is based on explicit generation of unitary operators using
Lie group decompositions.  Similar techniques have been applied to the problem of controlling
two-level systems \cite{PRA62n053409,IEEE39CDC1074} and especially particles with spin
\cite{dalessandro,qph0106115}.  Here we employ decompositions of the type discussed in 
\cite{PRA61n032106} to derive constructive control schemes for $N$-level systems.  We use
the rotating wave approximation (RWA) and require that \emph{each} allowed transition is 
\emph{selectively} addressable, for example by applying a field of the appropriate 
frequency, or by appropriate selection rules depending on the field polarization.  This 
means we must be able to ensure that each control pulse drives a single transition only, 
and that its effect on all other transitions is negligible.  These assumptions limit the
applicability of this approach to systems for which selective excitation of individual 
transitions is feasible such as atomic or molecular systems with well-separated transition
frequencies or particles in anharmonic potentials.  Certain other factors such as Doppler
shifts and inhomogeneous or homogeneous broadening must also be taken into account, and 
may require special consideration in specific circumstances.

However, for systems that satisfy the necessary conditions, the proposed technique has some
very attractive features.  It is constructive and can be used to solve a variety of control 
problems ranging from common problems with well-known solutions such as population transfer
between energy eigenstates to novel problems such as preparation of arbitrary superposition 
states or optimization of observables for $N$-level systems.  Moreover, although the control
schemes derived using this technique depend on the effective areas, and to a lesser extent, 
phases of the control pulses, the pulse \emph{shapes} are flexible, which implies that the 
control objective can be achieved using control pulses that are convenient from a practical
point of view such as square wave pulses with finite rise and decay times (SWP) or Gaussian
wavepackets (GWP).  SWP are a realistic approximation of bang-bang controls, which play an
important role in control theory and have been shown to be crucial for time-optimal control
\cite{JMP41p5262}.  Since both SWP and GWP can in principle be derived from continuous-wave
(CW) lasers using Pockel cells or other intensity modulating devices, this also opens the 
possibility for control of certain quantum systems using CW lasers, rather than more complex
pulsed laser systems and pulse-shaping techniques.

\section{Mathematical and physical framework}
\label{sec:basics}

We consider a non-dissipative quantum system with a discrete, finite energy spectrum such
as a generic $N$-level atom, molecule or particle in an (anharmonic) potential.  The free
evolution of the system is governed by the Schrodinger equation and determined by its 
internal Hamiltonian $\op{H}_0$, whose spectral representation is
\begin{equation} \label{eq:Hzero}
  \op{H}_0 = \sum_{n=1}^N E_n \ket{n}\bra{n},
\end{equation}
where $E_n$ are the energy levels and $\ket{n}$ the corresponding energy eigenstates of 
the system, which satisfy the stationary Schrodinger equation
\begin{equation} \label{eq:SSE}
   \op{H}_0 \ket{n} = E_n \ket{n}, \quad 1\le n\le N.
\end{equation}
Although this assumption is not required, we shall assume for simplicity that the energy 
levels $E_n$ are ordered in an increasing sequence, $E_1<E_2<\cdots<E_N$, where $N<\infty$
is the dimension of the Hilbert space of the system, and that the eigenstates $\{\ket{n}: 
n=1,\ldots,N\}$ form a complete orthonormal set.

The application of external control fields perturbs the system and gives rise to a new
Hamiltonian $\op{H}=\op{H}_0+\op{H}_I$, where $\op{H}_I$ is an interaction term.  If we
apply a field 
\begin{equation} \label{eq:fm}
  f_m(t) = 2 A_m(t) \cos(\omega_m t+\phi_m)
         = A_m(t) \left[e^{\rmi(\omega_m t+\phi_m)}+e^{-\rmi(\omega_m t+\phi_m)}\right]
\end{equation}
which is resonant with the frequency $\omega_m$ corresponding to the transition $\ket{m} 
\rightarrow \ket{m+1}$, and the pulse envelope $2A_m(t)$ is slowly varying with respect 
to the frequency $\omega_m$, then the rotating wave approximation (RWA) leads to the 
following interaction term
\begin{equation} \label{eq:Hm}
  \op{H}_m(f_m) =  A_m(t) d_m \left[e^{ \rmi(\omega_m t + \phi_m)} \ket{m}\bra{m+1}
                   + e^{-\rmi(\omega_m t + \phi_m)} \ket{m+1}\bra{m} \right]
\end{equation}
provided that (a) there are no other transitions with the same frequency $\omega_m$ and 
(b) off-resonant effects are negligible.  Note that the latter assumption is generally 
valid only if the Rabi frequency $\Omega_m$ of the driven transition is considerably less
than the minimum detuning from off-resonant transitions $\Delta\omega_{min}$, i.e.,
\begin{equation}  \label{eq:detuning}
  \max_t [\Omega_m(t)] = \max_t [2 A_m(t) d_m / \hbar] \ll \Delta\omega_{min},
\end{equation}
where $d_m$ is the dipole moment of the transition $\ket{m}\rightarrow\ket{m+1}$.  

The evolution of the controlled system is determined by the operator $\op{U}(t)$, which
satisfies the Schrodinger equation
\begin{equation} \label{eq:SE1}
  \rmi\hbar\frac{d}{dt}\op{U}(t) =
  \left\{ \op{H}_0 + \sum_{m=1}^M \op{H}_m[f_m(t)] \right\} \op{U}(t)
\end{equation}
and the initial condition $\op{U}(0)=\op{I}$, where $\op{I}$ is the identity operator.  

\section{Constructive control using Lie group decompositions}
\label{sec:Lie}

Our aim is to achieve a certain evolution of the system by applying a sequence of simple
control pulses.  Concretely, we seek to dynamically realize a desired unitary operator 
$\op{U}(t)$ at a certain target time $t=T$.  In some cases, we may not wish to specify a
target time in advance, in which case we attempt to achieve the control objective at some
later time $T>0$.

To solve the problem of finding the right sequence of control pulses, we apply the
interaction picture decomposition of the time-evolution operator $\op{U}(t)$,
\begin{equation} \label{eq:IPD}
  \op{U}(t) = \op{U}_0(t)\op{U}_I(t),
\end{equation}
where $\op{U}_0(t)$ is the time-evolution operator of the unperturbed system
\begin{equation} \label{eq:U0}
  \op{U}_0(t) = \exp\left( -\rmi\op{H}_0 t/\hbar \right)
              = \sum_{n=1}^N e^{-\rmi E_n t/\hbar} \ket{n}\bra{n}
\end{equation}
and $\op{U}_I(t)$ comprises the interaction with the control fields.  To obtain a 
dynamical law for the interaction operator $\op{U}_I(t)$, we note that inserting
\begin{eqnarray*}
  \rmi\hbar\frac{d}{dt}\op{U}(t)
  &=& \op{H}_0\op{U}_0(t)\op{U}_I(t) + \rmi\hbar\op{U}_0(t) \frac{d}{dt}\op{U}_I(t) \\
  \op{H}\op{U}(t) 
  &=& \op{H}_0\op{U}_0(t)\op{U}_I(t) + \sum_{m=1}^M \op{H}_m[f_m(t)]\op{U}_0(t)\op{U}_I(t)
\end{eqnarray*}
into the Schrodinger equation (\ref{eq:SE1}) gives
\begin{equation} \label{eq:SE2}
   \rmi\hbar\frac{d}{dt} \op{U}_I(t) 
 = \op{U}_0(t)^\dagger \left\{ \sum_{m=1}^M \op{H}_m[f_m(t)] \right\} 
   \op{U}_0(t) \op{U}_I(t).
\end{equation}
Applying (\ref{eq:U0}) and the rotating wave approximation Hamiltonian (\ref{eq:Hm}) to 
this equation leads after some simplification (see \ref{appendix:A}) to
\begin{equation}\label{eq:Omega}
  \frac{d}{dt}\op{U}_I(t)
 = \sum_{m=1}^M A_m(t) d_m/\hbar \left( \op{x}_m \sin\phi_m - \op{y}_m \cos\phi_m \right) 
   \op{U}_I(t)
\end{equation}
where we set $\op{e}_{m,n}= \ket{m}\bra{n}$ and define
\begin{equation}
  \op{x}_m    = \op{e}_{m,m+1} - \op{e}_{m+1,m}, \qquad
  \op{y}_m    = \rmi(\op{e}_{m,m+1} + \op{e}_{m+1,m}).
\end{equation}

Hence, if we apply a control pulse $f_k(t) = 2 A_k(t) \cos(\omega_m t+\phi_k)$ which is 
resonant with the transition frequency $\omega_m$ for a time period $t_{k-1}\le t\le t_k$
(and no other fields are applied during this time period) then we have
\begin{equation}
  \op{U}_I(t) = \op{V}_k(t)\op{U}_I(t_{k-1}),
\end{equation}
where the operator $\op{V}_k(t)$ is
\begin{equation} \label{eq:Vk}
  \op{V}_k(t)
  = \exp\left[ \frac{d_m}{\hbar} \int_{t_{k-1}}^t \!\!\! A_k(t') \, dt' 
    \left( \op{x}_m \sin\phi_k - \op{y}_m\cos\phi_k \right)\right].
\end{equation}
Thus, if we partition the time interval $[0,T]$ into $K$ subintervals $[t_{k-1},t_k]$ 
such that $t_0=0$ and $t_K=T$, and apply a sequence of non-overlapping control pulses, 
each resonant with one of the transition frequencies $\omega_m=\omega_{\sigma(k)}$, then
\begin{equation}
  \op{U}(T) = \op{U}_0(T)\op{U}_I(T)
            = e^{-\rmi\op{H}_0 T/\hbar}\op{V}_K \op{V}_{K-1} \cdots \op{V}_1,
\end{equation}
where the factors $\op{V}_k$ are
\begin{equation} \label{eq:Vk1}
  \op{V}_k = \exp\left[  \frac{d_{\sigma(k)}}{\hbar} \int_{t_{k-1}}^{t_k} \!\!\! A_k(t)\,dt
             \left(\op{x}_{\sigma(k)}\sin\phi_k-\op{y}_{\sigma(k)}\cos\phi_k \right)\right].
\end{equation}
$2 A_k(t)$ is the envelope of the $k$th pulse and $\sigma$ is a mapping from the index
set $\{1,\ldots,K\}$ to the control index set $\{1,\ldots, M\}$ that determines which 
of the control fields is active for $t \in [t_{k-1},t_k]$.

It has been shown \cite{PRA61n032106} that any unitary operator $\op{U}$ can be decomposed
into a product of operators of the type $\op{V}_k$ and a phase factor $e^{\rmi\Gamma}=
\det\op{U}$, i.e., there exists a positive real number $\Gamma$, real numbers $C_k$ and 
$\phi_k$ for $1\le k\le K$, and a mapping $\sigma$ from the index set $\{1,\ldots,K\}$ to
the control-sources index set $\{1,\ldots,M\}$ such that 
\begin{equation} \label{eq:Udecomp}
  \op{U}= e^{\rmi\Gamma}\op{V}_K\op{V}_{K-1}\cdots \op{V}_1,
\end{equation}
where the factors are
\begin{equation} \label{eq:Vk2}
  \op{V}_k = 
  \exp\left[C_k (\op{x}_{\sigma(k)}\sin\phi_k - \op{y}_{\sigma(k)} \cos\phi_k)\right].
\end{equation}
This decomposition of the target operator into a product of generators of the dynamical
Lie group determines the sequence in which the fields are to be turned on and off.  A 
general algorithm to determine the Lie group decomposition for an arbitrary operator 
$\op{U}$ is described in \ref{appendix:Udecomp}.

Note that in many cases the target operator $\op{U}$ is unique only up to phase factors,
i.e., two unitary operators $\op{U}_1$ and $\op{U}_2$ in $U(N)$ are equivalent if there 
exist values $\theta_n\in [0,2\pi]$ for $1\le n\le N$ such that
\begin{equation} \label{eq:Uequiv}
  \op{U}_2 = \op{U}_1 \left(\sum_{n=1}^N e^{\rmi\theta_n} \ket{n}\bra{n} \right)
\end{equation}
where $\ket{n}$ are the energy eigenstates.  For instance, if the initial state of the
system is an arbitrary ensemble of energy eigenstates
\begin{equation} \label{eq:rho0}
  \op{\rho}_0 = \sum_{n=1}^N w_n \ket{n}\bra{n},
\end{equation}
where $w_n$ is the initial population of state $\ket{n}$ satisfying $0\le w_n\le 1$ and 
$\sum_{n=1}^N w_n=1$, then we have
\[
 \op{U}_2\op{\rho}_0\op{U}_2^\dagger 
 = \op{U}_1 \left(\sum_{n=1}^N \ket{n} e^{\rmi\theta_n} w_n e^{-\rmi\theta_n} \bra{n}\right)
            \op{U}_1^\dagger 
 = \op{U}_1\op{\rho}_0\op{U}_1^\dagger
\]
i.e., the phase factors $e^{\rmi\theta_n}$ cancel.  Thus, if the initial state of the system
is an ensemble of energy eigenstates, which of course includes trivial ensembles such as 
pure energy eigenstates, then we only need to find a Lie group decomposition of the target
operator $\op{U}$ modulo phase factors, i.e., it suffices to find matrices $\op{V}_k$ such
that
\begin{equation}
  \op{U} \left(\sum_{n=1}^N e^{\rmi\theta_n} \ket{n}\bra{n} \right)
         = \op{V}_K \op{V}_{K-1} \cdots \op{V}_1.
\end{equation}
Note that decomposition modulo phase factors, when sufficient, is more efficient since
it requires in general up to $2(N-1)$ fewer steps than the general decomposition
algorithm.  See \ref{appendix:Udecomp} for details.

\section{Choice of pulse envelopes and pulse lengths}
\label{sec:amp}

Comparing equations (\ref{eq:Vk1}) and (\ref{eq:Vk2}) shows that 
\begin{equation} \label{eq:pulsearea}
   \frac{d_{\sigma(k)}}{\hbar} \int_{t_{k-1}}^{t_k} \!\!\! A_k(t) \, dt = C_k 
   \qquad \forall k,
\end{equation}
i.e., the effective pulse area of the $k$th pulse is $2C_k$ where $C_k$ is the constant in
decomposition (\ref{eq:Udecomp}).  However, the decomposition does not fix the pulse shapes,
i.e., we can choose pulse shapes that are convenient from a practical point of view such as
square wave pulses with finite rise and decay times (SWP) and Gaussian wavepackets (GWP), 
which can easily be produced in the laboratory.  For instance, in the optical regime both
SWP and GWP can be created using a combination of continuous-wave lasers and Pockel cells
or other intensity modulating devices.  Moreover, GWP are naturally derived from most pulsed
laser systems. 
\begin{figure}
\begin{center}
\begin{tabular}{ll}
(a) Square-wave pulse & (b) Gaussian pulse \\
\myincludegraphics[width=2in]{figures/pdf/SWP.pdf}{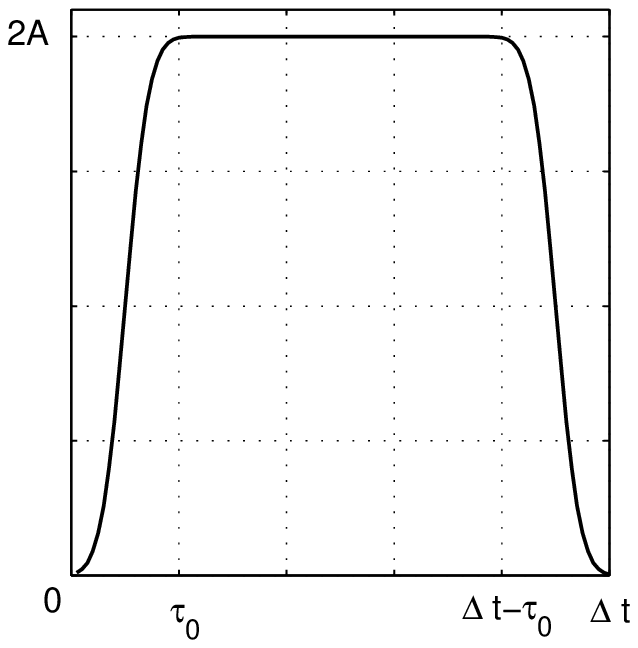} &
\myincludegraphics[width=2in]{figures/pdf/GWP.pdf}{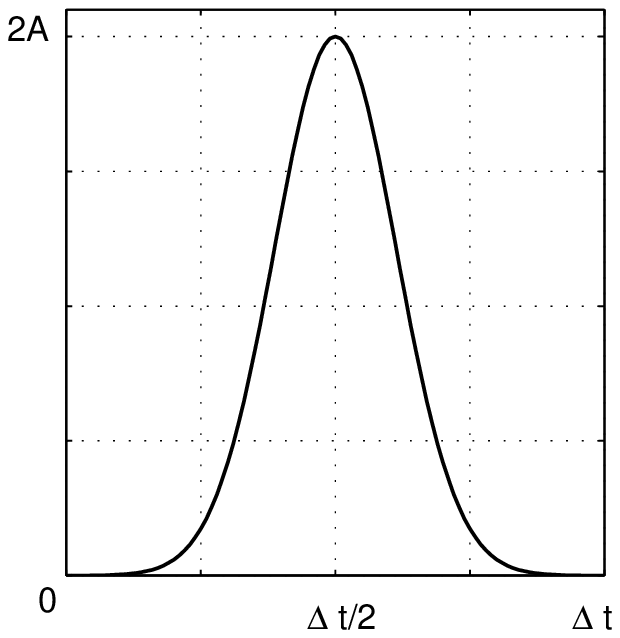} 
\end{tabular}
\end{center}
\caption{Square wave pulse of length $\Delta t_k$ with rise and decay time $\tau_0$ and
amplitude $2A_k$ (a) and Gaussian wavepacket with $q_k=\frac{4}{\Delta t_k}$ and peak 
amplitude $2A_k$ (b).} \label{Fig:pulses}
\end{figure}

\subsection{Square wave pulses}
\label{subsec:SWP}

The pulse area of an ideal square wave pulse of amplitude $2A_k$ and length $\Delta t_k$
is $2 A_k \Delta t_k$.  In order to accurately determine the pulse area of a realistic 
square wave pulse, however, we must take into account the finite rise and decay time 
$\tau_0$ of the pulse.  We can model the pulse envelopes of realistic SWP [see figure 
\ref{Fig:pulses} (a)] mathematically using
\begin{equation}
  2 A_k(t) = A_k\left\{2+\mbox{erf}\left[4(t-\tau_0/2)/\tau_0\right]
                 -\mbox{erf}\left[4(t-\Delta t+\tau_0/2)/\tau_0\right]\right\} 
\end{equation}
where $\mbox{erf}(x)$ is the error function
\[
   \mbox{erf}(x) = \frac{2}{\sqrt{\pi}} \int_0^x \!\!\! e^{-t^2} \, dt.
\]
Although this envelope function may appear complicated, it can easily be checked that the
area bounded by this function and $t_{k-1}\le t\le t_k$ equals the area of a rectangle of
width $\Delta t_k - \tau_0$ and height $2A_k$.  Thus, the pulse area $\int_{\Delta t_k} 
2A_k(t)\,dt$ of a realistic square wave pulse is $2A_k(\Delta t_k-\tau_0)$, and equation
(\ref{eq:pulsearea}) shows that the amplitude of the pulse is determined by 
\begin{equation}\label{eq:Ak1}
   A_k = \frac{1}{\Delta t_k-\tau_0} \times \frac{\hbar}{d_{\sigma(k)}} \times C_k 
       = \frac{\hbar C_k}{(\Delta t_k -\tau_0) d_{\sigma(k)}},
\end{equation}
where $d_{\sigma(k)}$ is the dipole moment of the driven transition.  

To ensure selective excitation, the contribution of Fourier components with $\Delta\omega 
\ge\Delta\omega_{min}$ must be negligeable.  Noting that the Fourier transform of an ideal
SWP ($\tau_0=0$) of length $\Delta t_k$ and amplitude $2A_k$ is
\begin{equation} 
  F(\Delta\omega) = 2 A_k \sqrt{\frac{2}{\pi}} 
                    \frac{\sin(\frac{1}{2}\Delta t_k\Delta\omega)}{\Delta\omega},
\end{equation}
where $\Delta\omega$ is the detuning from the pulse frequency $\omega_m$, shows that $F(0)
=\sqrt{\frac{2}{\pi}}A_k\Delta t_k$ and 
\[
  \frac{F(\Delta\omega)}{F(0)} =
  \frac{\sin(\frac{1}{2}\Delta t_k\Delta\omega)}{\frac{1}{2}\Delta t_k\Delta\omega}, 
\]
i.e., $\frac{F(\Delta\omega)}{F(0)} \ll 1$ if $\Delta t_k\Delta\omega\gg 1$.  Thus, 
contributions from Fourier components with $\Delta\omega \ge \Delta\omega_{min}$ will
be negligible if $\Delta t_k\gg\Delta\omega_{min}^{-1}$.

Furthermore, noting that $C_k\le\frac{\pi}{2}$, the peak Rabi frequency for a square wave
pulse of length $\Delta t_k$ with rise and decay time $\tau_0$ is
\begin{equation} \label{eq:Rabi1}
  \max_{t_{k-1} \le t \le t_k} \left[ 2 A_k(t) d_{\sigma(k)}/\hbar \right]
                 = \frac{2 C_k}{\Delta t_k-\tau_0}
                 \le \frac{\pi}{\Delta t_k-\tau_0}.
\end{equation}
Hence, the Rabi frequency and the amplitude of the pulse can be adjusted by changing
the pulse length $\Delta t_k$, which allows us to ensure that (\ref{eq:detuning}) is 
satisfied, and enforce laboratory constraints on the strengths of the control fields.  

We can also give an estimate of the time required to implement arbitrary unitary operators
given certain bounds on the field strength.  If the maximum strength of the field produced 
by the $m$th laser is $A_{m,max}$, i.e, $f_m(t)=2 A_m(t)\cos(\omega_m t+\phi_m)\le A_{m,max}$
then the time required to perform a rotation by $C_k$ on the transition $\ket{m}\rightarrow
\ket{m+1}$ using a SWP with rise and decay time $\tau_0$ is  
\begin{equation} \label{eq:tmax:SWP}
  \Delta t_m^{SWP} = \frac{2 C_k\hbar}{A_{m,max} d_m} + \tau_0 
                   \le \frac{\pi\hbar}{A_{m,max} d_m} + \tau_0.
\end{equation}
\ref{appendix:Udecomp} shows that any unitary operator $\op{U}$ can be generated up to 
equivalence (\ref{eq:Uequiv}) by performing at most $N-m$ rotations by $C\le\frac{\pi}{2}$ 
on each transition $\ket{m}\rightarrow\ket{m+1}$ for $m=1,2,\dots, N-1$.  Hence, any unitary
operator can be implemented up to equivalence using SWP of amplitude $A_{m,max}$ in at most
time $T$, where
\begin{equation}
  T = \sum_{m=1}^{N-1} \max(\Delta t_m^{SWP}) (N-m)
    = \sum_{m=1}^{N-1} \left(\frac{\pi\hbar}{A_{m,max}d_m} + \tau_0\right) (N-m).
\end{equation}
Since two additional rotations on each transition are required to generate $\op{U}$ exactly,
the latter can be accomplished in time $T'\ge\sum_{m=1}^{N-1}\max(\Delta t_m^{SWP})(N-m+2)$.

\subsection{Gaussian wavepackets}
\label{subsec:GWP}

To model a Gaussian wavepacket [see figure \ref{Fig:pulses} (b)] of peak amplitude $2A_k$
centered at $t_k^*=t_{k-1}+\frac{1}{2}\Delta t_k$, we choose the pulse envelope
\begin{equation}
  2A_k(t) = 2A_k \exp\left[-q_k^2 (t-\Delta t_k/2 - t_{k-1})^2 \right].
\end{equation}
The constant $q_k$ determines the width of the wavepacket.  The pulse area of a Gaussian
wavepacket is $\sqrt{\pi}/q_k$ provided that the time interval $\Delta t_k$ is large enough 
to justify the assumption
\[
  \int_{t_{k-1}}^{t_k} \exp\left[-q_k^2 (t-\Delta t_k/2 - t_{k-1})^2 \right]\, dt
  \approx \int_{-\infty}^{+\infty} e^{-q^2 \tau^2} \, d\tau 
  = \frac{\sqrt{\pi}}{q_k}.
\]
In the following we choose $q_k = 4/\Delta t_k$, which guarantees that over 99\% of the 
$k$th pulse is contained in the control interval $[t_{k-1},t_k]$ since
\[
 \int_{-\Delta t_k/2}^{\Delta t_k/2} e^{-q_k^2 t^2}\, dt
 = \frac{\sqrt{\pi}}{q_k} \mbox{erf}(q_k \Delta t_k/2)
\]
and $\mbox{erf}(2)=0.995322$.  Thus, (\ref{eq:pulsearea}) shows that the peak amplitude
$2A_k$ of the GWP is determined by
\begin{equation} \label{eq:Ak2}
  A_k = \frac{q_k}{\sqrt{\pi}} \times \frac{\hbar}{d_{\sigma(k)}} \times C_k 
      = \frac{4\hbar C_k}{\sqrt{\pi} \Delta t_k d_{\sigma(k)}}.
\end{equation}

Again, to ensure selective excitation, the contribution of Fourier components with $\Delta
\omega \ge\Delta\omega_{min}$ must be negligeable.  Noting that the Fourier transform of a 
Gaussian wavepacket with $q_k=4/\Delta t_k$ and amplitude $2A_k$ is
\begin{equation} 
  F(\Delta\omega) = \frac{2 A_k}{\sqrt{2}q_k} 
                    \exp \left[ -\frac{\Delta\omega^2}{4q_k^2} \right]
                  = \frac{\Delta t_k A_k}{2\sqrt{2}} 
                    \exp ( -\Delta\omega^2\Delta t_k^2/16) 
\end{equation}
where $\Delta\omega$ is the detuning from the pulse frequency $\omega_m$, shows that 
\[
  \frac{F(\Delta\omega)}{F(0)} = \exp( -\Delta\omega^2\Delta t_k^2/16) 
\]
i.e., $\frac{F(\Delta\omega)}{F(0)} \ll 1$ if $\Delta t_k\Delta\omega\gg 4$.  Thus, 
contributions from Fourier components with $\Delta\omega \ge \Delta\omega_{min}$ will
be negligible if $\Delta t_k\gg 4\Delta\omega_{min}^{-1}$.

Furthermore, noting that $C_k\le\frac{\pi}{2}$, the peak Rabi frequency for a Gaussian
pulse of length $\Delta t_k$ with $q_k=4/\Delta t_k$ is 
\begin{equation} \label{eq:Rabi2}
  \max_{t_{k-1} \le t \le t-k} \left[ 2 A_k(t) d_{\sigma(k)}/\hbar \right]
                 = \frac{8 C_k}{\sqrt{\pi}\Delta t_k}
                 \le \frac{4\sqrt{\pi}}{\Delta t_k}.
\end{equation}
Hence, the Rabi frequency can again be adjusted by changing the pulse length $\Delta t_k$,
which allows us to ensure that (\ref{eq:detuning}) is satisfied and enforce laboratory 
constraints on the strengths of the control fields.  

Again, we can give an estimate of the time required to implement arbitrary unitary operators
given certain bounds on the field strength.  If the maximum strength of the field produced
by the $m$th laser is $A_{m,max}$, i.e, $f_m(t)=2 A_m(t)\cos(\omega_m t+\phi_m)\le A_{m,max}$
then the time required to perform a rotation by $C_k$ on the transition $\ket{m}\rightarrow
\ket{m+1}$ using GWP with $q_k=4/\Delta t_k$ is  
\begin{equation} \label{eq:tmax:GWP}
  \Delta t_m^{GWP} = \frac{8 C_k\hbar}{\sqrt{\pi} A_{m,max} d_m}
                   \le \frac{4\sqrt{\pi}\hbar}{A_{m,max} d_m}.
\end{equation}
Since any unitary operator $\op{U}$ can be generated up to equivalence (\ref{eq:Uequiv}) by 
performing at most $N-m$ rotations by $C_k \le \frac{\pi}{2}$ on each transition $\ket{m}
\rightarrow\ket{m+1}$ for $m=1,2,\dots, N-1$, the time required to implement $\op{U}$ up to
equivalence using GWP of (peak) amplitude $A_{m,max}$ is at most 
\begin{equation}
  T = \sum_{m=1}^{N-1} \max(\Delta t_m^{GWP}) (N-m)
    = \sum_{m=1}^{N-1} \left(\frac{4\sqrt{\pi}\hbar}{A_{m,max}d_m}\right) (N-m).
\end{equation}
Since two additional rotations on each transition are required to generate $\op{U}$ exactly,
the latter can be accomplished in time $T'\ge\sum_{m=1}^{N-1}\max(\Delta t_m^{GWP})(N-m+2)$.

\section{Physical systems used for illustration}
\label{sec:examples}

In the following sections we shall apply these results to various control problems.  For
numerical illustrations of our control schemes, we shall consider 
\begin{enumerate}
\item a four-level model of the electronic states of Rubidium (\Rb87)
\item a four-level Morse oscillator model of the vibrational modes of hydrogen fluoride.
\end{enumerate}

\begin{figure}
\begin{center}
\begin{tabular}{l@{}l}
(a) \Rb87 & (b) HF \\
\myincludegraphics[width=3in]{figures/pdf/Rb4.pdf}{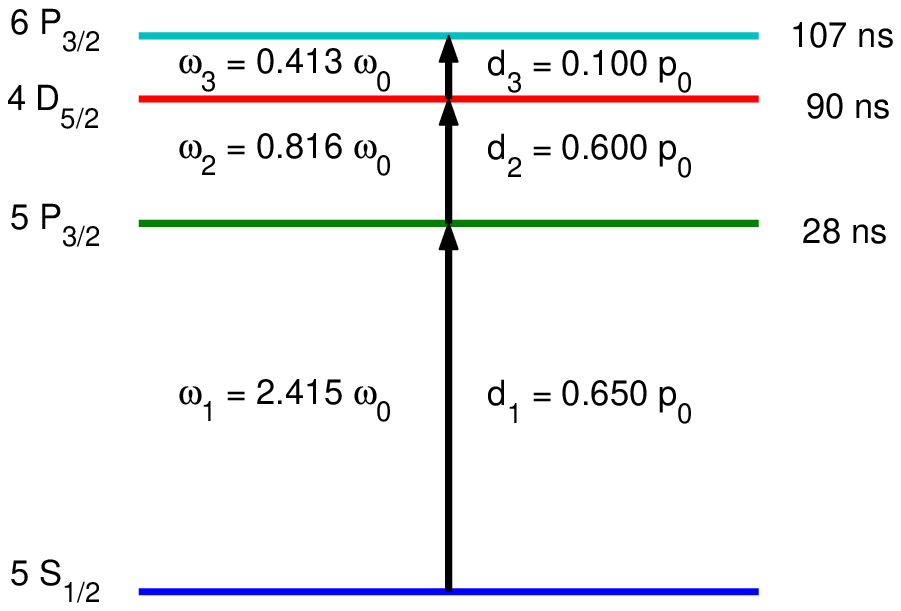} &
\myincludegraphics[width=3in]{figures/pdf/HF4.pdf}{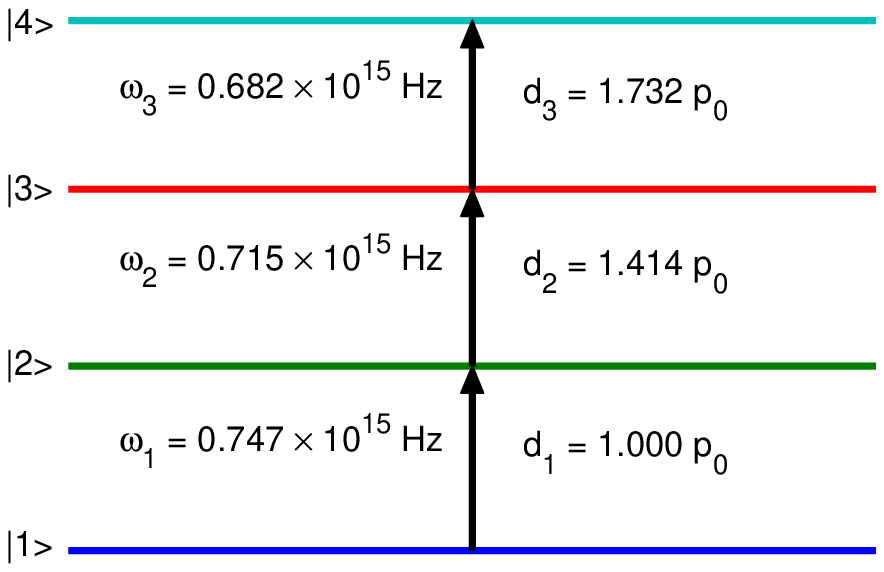}
\end{tabular}
\caption{Transition diagram for Rubidium (a) and hydrogen fluoride (b).  For \Rb87 the
constant $\omega_0=10^{15}$ Hz and the electric dipole moment unit is $p_0=4.89 \times 
10^{-29}$ C m.  For HF the electric dipole moment unit is $p_0=3.24 \times 10^{-31}$ C m.}
\label{fig:system} 
\end{center}
\end{figure}

For Rubidium (\Rb87) we consider four electronic states, which we label as follows: 
$\ket{1}=\ket{5 S_{1/2}}$, $\ket{2}=\ket{5 P_{3/2}}$, $\ket{3}=\ket{4 D_{1/2}}$ and 
$\ket{4}=\ket{6 P_{3/2}}$, where $\ket{1}$ is the ground state.  Figure \ref{fig:system}
(a) shows the coupling diagram with transition frequencies and dipole moments.  

For hydrogen fluoride (HF) we use the Morse oscillator model given in \cite{PRL65p2355}. The 
energy levels corresponding to the vibrational states $\ket{n}$ are 
\[
  E_n = \hbar\omega_0 \, (n - \mbox{$\frac{1}{2}$})
        \left[1 - \mbox{$\frac{B}{2}$}(n - \mbox{$\frac{1}{2}$})\right]
\] 
where $\omega_0=0.78 \times 10^{15}$ Hz and $B=0.0419$.  The frequencies for transitions 
between adjacent energy levels are $\omega_n=\hbar\omega_0(1-B n)$ and the corresponding
transition dipole moments are $d_n=p_0\sqrt{n}$ with $p_0=3.24\times 10^{-31}$ C m, which
leads to the values shown in figure \ref{fig:system} (b).  Although there are 24 bound 
vibrational states for this model, we only consider the four lowest vibrational modes $n 
= 1,2,3,4$, where $\ket{1}$ is the ground state.  

Since we have made several approximations in developing our control approach using Lie 
group decompositions, we must ensure that the assumptions we made are valid for the systems
we consider:  
\begin{enumerate}
\item No two transitions have the same transition frequency.%
      \footnote{Assumption (\ref{hyp:a}) can be relaxed if we can distinguish 
      transitions with the same transition frequency by other means, e.g., by 
      using fields with different polarizations.}
      \label{hyp:a}
\item Dissipative effects are negligible. \label{hyp:b}
\item The effect of the pulse on off-resonant transitions is negligible. \label{hyp:c}
\end{enumerate}

Note that both models satisfy hypothesis (\ref{hyp:a}).  Furthermore, the main source of 
dissipation for both systems is spontaneous emission.  Thus, dissipative effects will be
negligible provided that the control pulses are much shorter than the lifetimes of the 
excited states.  Since the lifetimes of the excited electronic states for \Rb87 are $28$,
$90$ and $107$ ns, respectively, hypothesis (\ref{hyp:b}) will be satisfied for control 
pulses in the sub-nanosecond regime.  Similarly for HF.

Hypothesis (\ref{hyp:c}) will be satisfied provided that:
\begin{enumerate}
\item[(a)] 
     The Fourier spectrum of the pulse does not overlap with other transition frequencies,
     i.e., the frequency dispersion of the pulse is less than the minimum detuning from 
     off-resonant transitions.
\item[(b)] 
     Equation (\ref{eq:detuning}) holds, i.e., the Rabi frequency of each driven transition
     is much smaller than the minimum detuning from off-resonant transitions. 
\end{enumerate}
Since the minimum detuning from off-resonant transitions is $\Delta\omega_{min}\approx 4
\times 10^{14}$ Hz for \Rb87 and $\Delta\omega_{min}\approx 3.27\times 10^{13}$ Hz for HF, 
the pulse length $\Delta t_k$ should be at least $10^{-12}$ and $10^{-11}$ seconds, 
respectively, to ensure that the frequency dispersion of the pulse is sufficiently small.  
Moreover, inserting the values for $\Delta\omega_{min}$ as well as (\ref{eq:Rabi1}) and 
(\ref{eq:Rabi2}), respectively, into equation (\ref{eq:detuning}) shows again that we must
choose the pulse lengths such that $\Delta t_k\gg 10^{-14}$ s for \Rb87 and $\Delta t_k\gg
10^{-13}$ s for HF to ensure that the second condition above is met.  In the following, we
shall choose $\Delta t_k=2\times 10^{-10}$ seconds (200 ps) for all pulses, which ensures
that both hypotheses (\ref{hyp:b}) and (\ref{hyp:c}) are met for both \Rb87 and HF.  
Moreover, such pulses are also experimentally realizable. 

Note that the energy levels for \Rb87 are multiply degenerate due to hyperfine and other
effects.  Since the detuning between the $F=1$ and $F=2$ sublevels of the $5 S_{1/2}$ ground
state is rather large (6.8 GHz), we may wish to be precise and choose $\ket{1}=\ket{5S_{1/2},
F=1}$, for instance, but we shall generally ignore the hyperfine energy level structure here.
For the cases we consider in this paper, this is justified since the frequency differences
between the hyperfine levels (except for the ground state) are on the order of several 
hundred MHz or less, which corresponds to detunings of $\Delta\omega\le 10^8$ Hz, which we
cannot resolve with 200 ps pulses for reasons outlined above.

\section{Population transfer $\ket{1}\rightarrow\ket{N}$ for a $N$-level system}
\label{sec:poptransfer}

We shall first apply the decomposition technique described above to the rather elementary 
control problem of population transfer between energy eigenstates to better illustrate 
the technique.  Concretely, we consider the problem of transferring the population of the 
ground state $\ket{1}$ to the excited state $\ket{N}$ by applying a sequence of control 
pulses, each resonant with one of the transitions frequencies $\omega_m$.  It can easily
be verified that any evolution operator $\op{U}$ of the form
\begin{equation} \label{eq:U1}
  \op{U} = \left( \begin{array}{c|c}
            \vec{0}       & \; A_{N-1} \\\hline
            e^{\rmi\theta}\; & \; \vec{0}
           \end{array} \right),
\end{equation}
where $A_{N-1}$ is an arbitrary unitary $(N-1)\times (N-1)$ matrix, $e^{\rmi\theta}$ is 
an arbitrary phase factor and $\vec{0}$ is a vector whose $N-1$ elements are $0$, achieves
the control objective since
\[
  \left(\begin{array}{c|c}
            \vec{0}       & \; A_{N-1} \\\hline
            e^{\rmi\theta}\; & \; \vec{0}
          \end{array}\right) 
   \left( \begin{array}{c} 1       \\ \vec{0}       \end{array} \right)
 = \left( \begin{array}{c} \vec{0} \\ e^{\rmi\theta_N} \end{array} \right)
\]
and thus the population of state $\ket{N}$ is equal to $\sqrt{e^{-\rmi\theta_N}
e^{\rmi\theta_N}}=1$ after application of $\op{U}$.  Next, we observe that setting 
\begin{equation}\label{eq:Udecomp1}
   \op{U}  = \op{U}_0(T)\op{U}_I, \quad
   \op{U}_I= \op{V}_{N-1} \op{V}_{N-2} \cdots \op{V}_1,
\end{equation}
where the factors are
\begin{eqnarray} \label{eq:Vm}
 \op{V}_m &=& \exp\left[ \frac{\pi}{2}
              \left(\op{x}_m\sin\phi_m-\op{y}_m\cos\phi_m\right)\right] \\
          &=& -\rmi(e^{\rmi\phi_m} \op{e}_{m,m+1}+e^{-\rmi\phi_m}\op{e}_{m+1,m})
               + \sum_{n\neq m,m+1} \op{e}_{n,n} \nonumber
\end{eqnarray}
for $1\le m\le N-1$, always leads to a $\op{U}$ of the form (\ref{eq:U1}), independent 
of the initial pulse phases $\phi_m$.  

The factorization (\ref{eq:Udecomp1}) corresponds to a sequence of $N-1$ control pulses 
in which the $m$th pulse is resonant with the frequency $\omega_m$ of the transition 
$\ket{m} \rightarrow \ket{m+1}$ and has effective pulse area $\pi$.  Thus, the solution 
obtained using the decomposition technique is an intuitive sequence of $\pi$-pulses 
designed to transfer the population step by step to the target level.  

\begin{figure}
\begin{center}
\begin{tabular}{l@{}l}
(a) Square wave pulses & (b) Gaussian pulses \\
\myincludegraphics[width=3in]{figures/pdf/Rb4_PopTransfer_SWP.pdf}{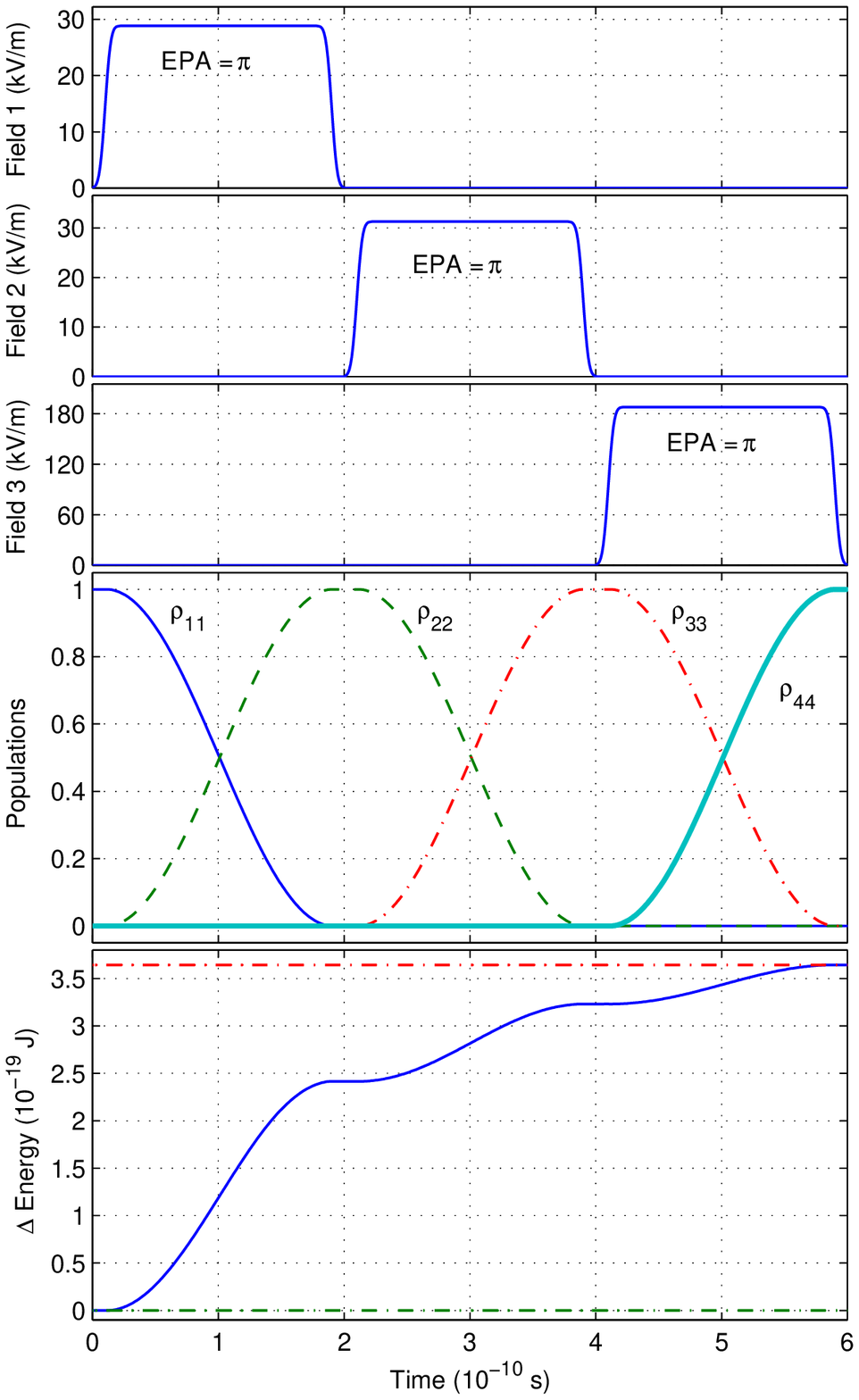} &
\myincludegraphics[width=3in]{figures/pdf/Rb4_PopTransfer_GWP.pdf}{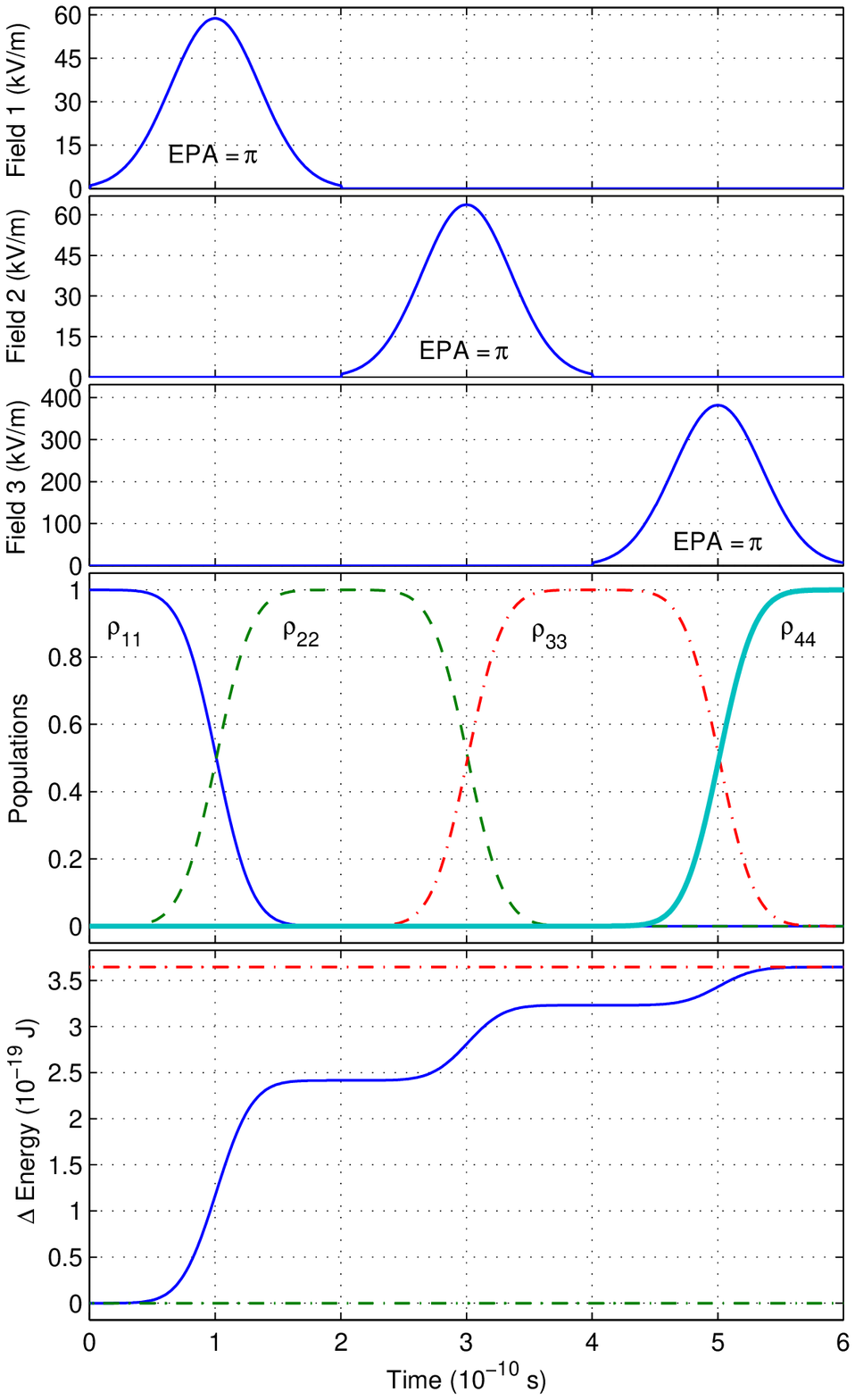}
\end{tabular}
\end{center}
\caption{Population transfer from the ground state $\ket{1}=\ket{5S_{1/2}}$ to the excited
state $\ket{4}=\ket{6P_{3/2}}$ for \Rb87 using (a) three 200 ps square wave pulses with 
rise and decay time $\tau_0=20$ ps and (b) three 200 ps Gaussian pulses with $q=2\times
10^{10}$ Hz.  The top graphs show the pulse envelopes $A_k(t)$.  The effective pulse area
$\mbox{EPA}=\int_{\Delta t_m} 2 A_m(t) d_m\, dt$ of all pulses is $\pi$.  The labels `Field
$m$' indicate that the corresponding pulses are resonant with the frequency $\omega_m$ of 
the transition $\ket{m} \rightarrow \ket{m+1}$.}
\label{Fig:PopTransfer} 
\end{figure}

The results of illustrative computations for the four-level \Rb87 system introduced above
are shown in figure \ref{Fig:PopTransfer}.  The top graphs show the pulse sequence for 
square wave pulses (a) and Gaussian control pulses (b).  The corresponding evolution of 
the energy-level populations shows that the populations of the intermediate levels increase
and decrease intermittently as expected, while the population of target level $\ket{4}$ 
reaches one at the final time.  The bottom graph shows that the energy of the system 
increases monotonically from its kinematical minimum value at $t=0$ to its maximum value
at the final time as predicted.  The basic response of the system is the same for square
wave pulses and Gaussian pulses.  However, the energy increases more uniformly for square
wave pulses, while Gaussian pulses tend to result in short, steep increases with long 
intermittent plateau regions.  Square wave pulses may therefore be a better choice if one
wishes to minimize the time the system spends in intermediate states with short lifetimes.
Gaussian wavepackets, on the other hand, have the advantage of minimal frequency dispersion
and are thus less likely to induce unwanted off-resonant effects.

As regards the field strengths, note that for 200 ps pulses up to 380 kV/m are required 
for SWP, and up to 780 kV/m for Gaussian pulses, which corresponds to (peak) intensities 
$I=\epsilon_0 c E^2$ of up to $40 \mbox{ kW/cm}^2$ (SWP) and $160 \mbox{ kW/cm}^2$ (GWP),
respectively.  Achieving these intensities experimentally with CW lasers is feasible using
a combination of sufficiently powerful lasers and beam focusing techniques.  Since pulsed
laser systems with 1 mJ output for picosecond pulses are common, intensities of up to $10^7
\mbox{ W/cm}^2$ should be easy to achieve for these systems.

Note that we chose pulses of fixed length $200$ ps and allowed the pulse amplitudes to 
vary.  Had we instead fixed the strength of the fields to be $2 A_k = 10^5$ V/m, say, then
the length $\Delta t_k$ of the control pulses according to (\ref{eq:tmax:SWP}) would have 
been 124.2, 132.7 and 697.1 ps, respectively, for SWP with $\tau_0=20$ ps.  For Gaussian 
pulses with $q_k=4/\Delta t_k$, the pulse length according to (\ref{eq:tmax:GWP}) would 
have been 235.1, 254.7 and 1528.2 ps, respectively.  Thus, instead of 600 ps in both cases,
the time required to achieve the control objective would have been 954 ps for SWP and 2018
ps for GWP.

\section{Inversion of ensemble populations for a mixed-state system}
\label{sec:inversion}

Sequences of $\pi$-pulses similar to the ones derived in the previous section have played
an important role in the theory of atomic excitation \cite{90Shore} and have been applied 
to the problem of vibrational excitation of molecules in both theory \cite{CP267p173} and
experiment \cite{CPL270p45}.  The decomposition technique is an important tool since it 
allows us to generalize the intuitive control schemes for population transfer between 
energy eigenstates to obtain similar schemes for a variety of more complicated problems,
as we shall demonstrate now.  

The first example we consider is a generalization of the population transfer problem to 
mixed-state systems.  The objective is to achieve a complete inversion of the ensemble 
populations given an arbitrary initial state of the form (\ref{eq:rho0}).  This control
operation can be regarded as an ensemble--NOT gate for mixed-state systems, not to be 
confused with other NOT--gates such as the U--NOT gate \cite{JMO47p211}.  Complete 
inversion of the ensemble populations requires an evolution operator
\begin{equation} \label{eq:U2}
  \op{U}=\left(\begin{array}{cccccc}
            0 & 0 & \cdots & 0 & e^{\rmi\theta_1} \\
            0 & 0 & \cdots & e^{\rmi\theta_2} & 0 \\
            \vdots& \vdots & & \vdots & \vdots \\
            0 & e^{\rmi\theta_{N-1}} & \cdots & 0 & 0\\
            e^{\rmi\theta_N} & 0     & \cdots & 0 & 0
          \end{array}\right),
\end{equation}
where the $e^{\rmi\theta_n}$ are arbitrary phase factors.  Assuming as before that each 
transition between adjacent energy levels can be individually addressed, the generators
of the dynamical Lie algebra are again of the form (\ref{eq:Vk1}) and the target operator
(\ref{eq:U2}) can be written as a product of these generators
\begin{equation} \label{eq:Udecomp2}
  \op{U} = \op{U}_0(T) \prod_{\ell=N-1}^{1} \left[\prod_{m=1}^\ell \op{V}_m \right],
\end{equation}
where the factors $\op{V}_m$ are as defined in (\ref{eq:Vm}).  The decomposition 
(\ref{eq:Udecomp2}) corresponds to a sequence of $K=N(N-1)/2$ pulses in which the $k$th
pulse is resonant with the transition $\ket{\sigma(k)}\rightarrow \ket{\sigma(k)+1}$ and
has effective pulse area $\pi$, where 
\[
  \sigma([1,\ldots,K]) = 
  [1, 2, \cdots, N-1; 1, 2 \cdots N-2; 1, 2, \cdots, N-3; \cdots; 1, 2; 1].
\]
This pulse scheme does \emph{not} depend on the values of the initial populations, i.e., 
a complete inversion of the ensemble populations is achieved for any initial ensemble.  
Moreover, if the initial populations are mutually distinct, i.e., $w_n\neq w_m$ for $n
\neq m$, then the decomposition is optimal in the sense that a complete inversion of the
ensemble populations cannot be achieved with fewer than $K$ control pulses.

To illustrate the control scheme, let us apply it to the four-level Morse oscillator 
model for the vibrational modes of HF discussed above.  For the purpose of the computer
simulations, we randomly choose the initial populations to be $w_1=0.4$, $w_2=0.3$, 
$w_3=0.2$ and $w_4=0.1$, but recall that any initial ensemble would do, i.e., we could
have chosen a thermal ensemble given by a Boltzmann distribution or another ensemble 
instead.  Our goal is to create an ensemble where the populations of the energy 
eigenstates are reversed, i.e., where $\ket{1}$ has population $w_4$, $\ket{2}$ has
population $w_3$, $\ket{3}$ has population $w_2$, and $\ket{4}$ has population $w_1$.

\begin{figure}
\begin{center}
\begin{tabular}{l@{}l}
(a) Square wave pulses & (b) Gaussian pulses \\
\myincludegraphics[width=3in]{figures/pdf/HF4_PopInversion_SWP.pdf}{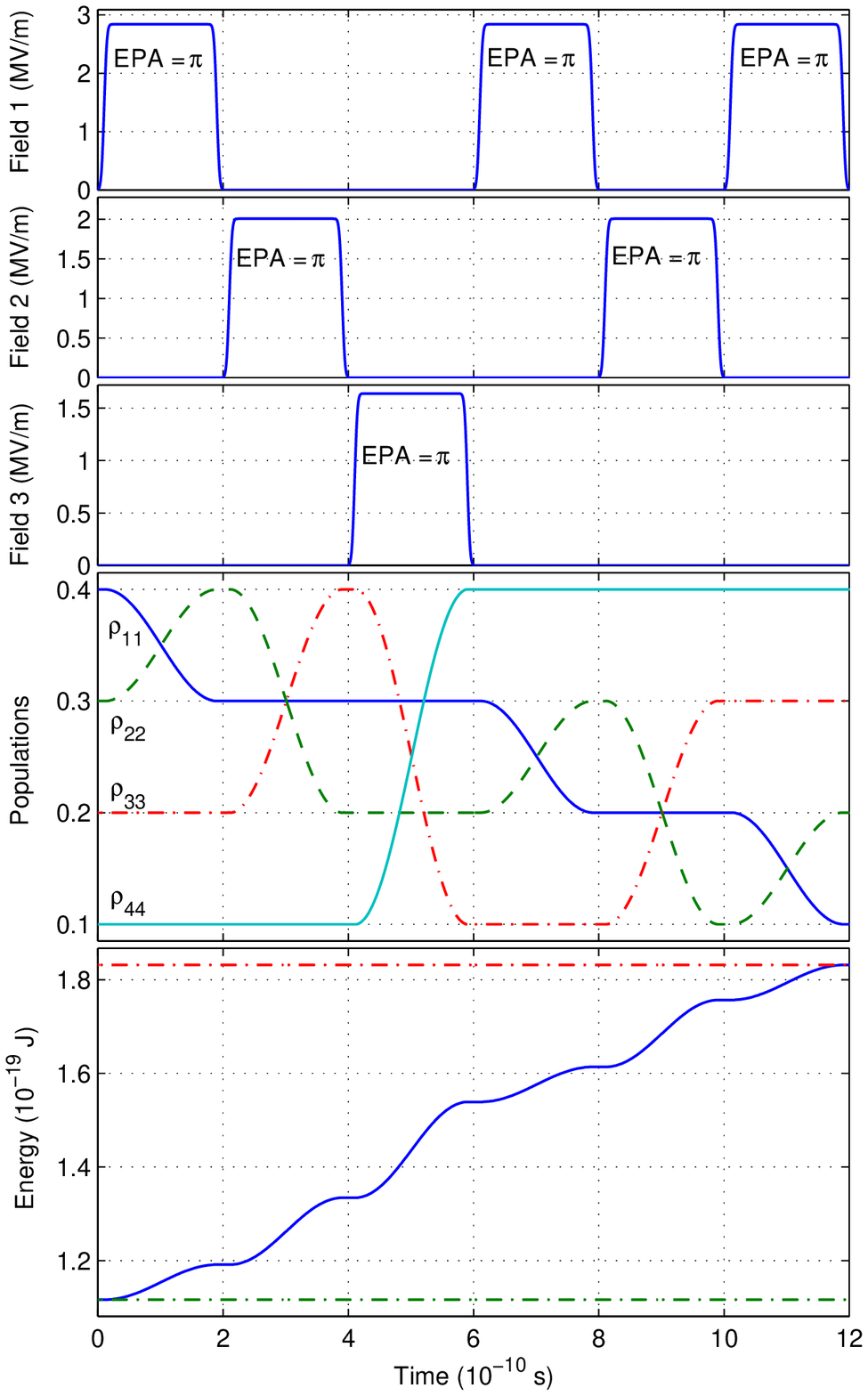} 
&
\myincludegraphics[width=3in]{figures/pdf/HF4_PopInversion_GWP.pdf}{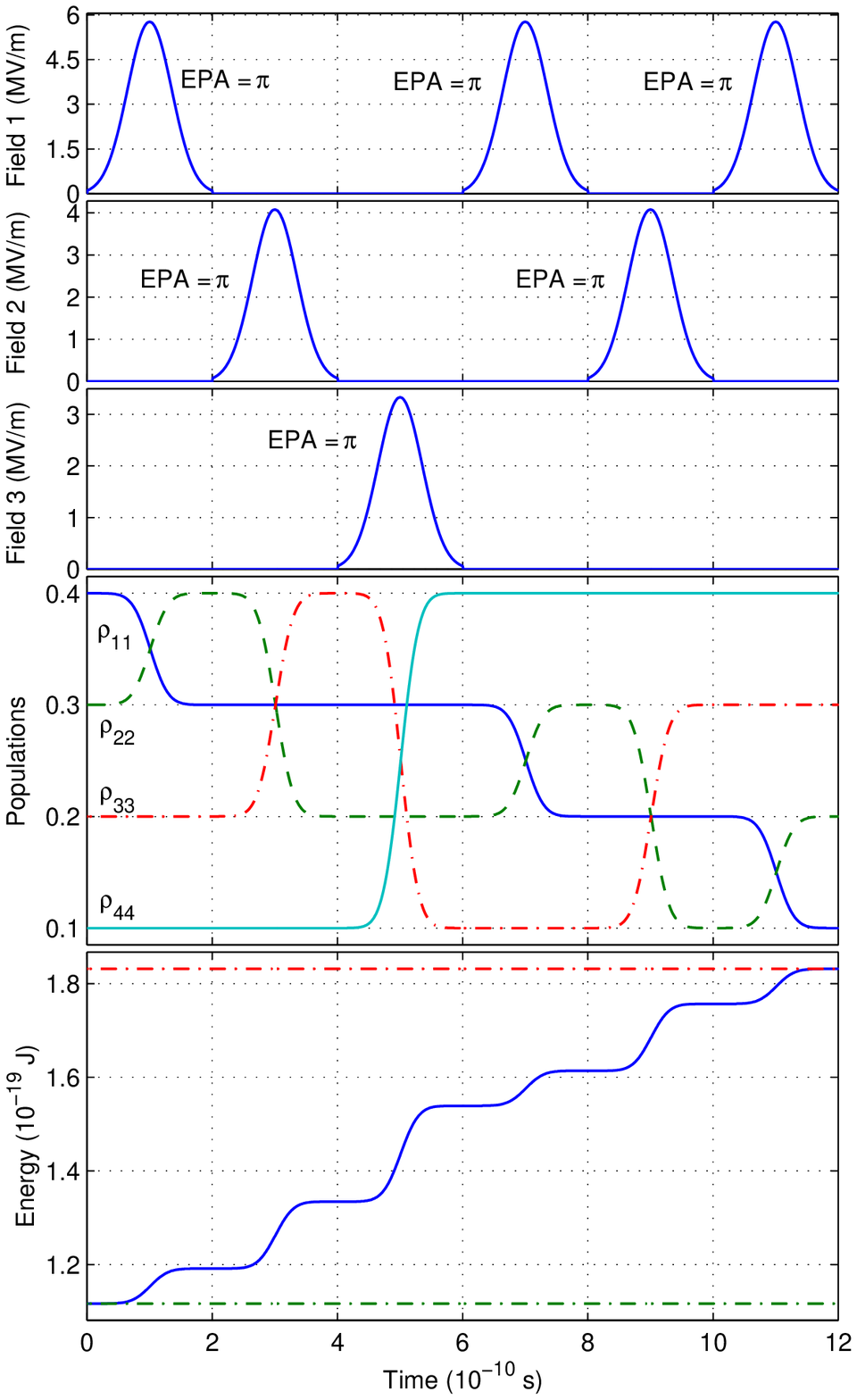}
\end{tabular}
\end{center}
\caption{Inversion of the ensemble populations for the vibrational modes of HF using 
(a) six square wave control pulses with rise and decay time $\tau_0=20$ ps, and (b) six 
Gaussian pulses with $q=2\times 10^{10}$ Hz.  The top graphs show the pulse envelopes 
$A_k(t)$. The effective pulse area $\mbox{EPA}=\int_{\Delta t_m} 2 A_m(t) d_m\, dt$ of 
all pulses is $\pi$.  The labels `Field $m$' indicate that the corresponding pulses are 
resonant with the frequency $\omega_m$ of the transition $\ket{m} \rightarrow\ket{m+1}$.}
\label{Fig:PopInversion} 
\end{figure}

Figure \ref{Fig:PopInversion} shows the results of control simulations using square wave 
and Gaussian control pulses, respectively.  Note that each pulse in the control sequence 
interchanges the populations of two adjacent energy levels until a complete inversion of
the populations is achieved.  For our four-level system the effect of the controls on 
the populations can be summarized as follows
\begin{center}
\setlength{\unitlength}{3000sp}%
\begingroup\makeatletter\ifx\SetFigFont\undefined%
\gdef\SetFigFont#1#2#3#4#5{%
  \reset@font\fontsize{#1}{#2pt}%
  \fontfamily{#3}\fontseries{#4}\fontshape{#5}%
  \selectfont}%
\fi\endgroup%
\begin{picture}(3604,1755)(601,-1261)
\thinlines
\put(880,-166){\vector( 1, 1){285}}
\put(880, 52){\vector( 1,-1){285}}
\put(1480,-646){\vector( 1, 1){285}}
\put(1480,-428){\vector( 1,-1){285}}
\put(2080,-1096){\vector( 1, 1){285}}
\put(2080,-878){\vector( 1,-1){285}}
\put(2680,-166){\vector( 1, 1){285}}
\put(2680, 52){\vector( 1,-1){285}}
\put(3280,-661){\vector( 1, 1){285}}
\put(3280,-443){\vector( 1,-1){285}}
\put(3880,-196){\vector( 1, 1){285}}
\put(3880, 22){\vector( 1,-1){285}}
\put(1,90){\makebox(0,0)[lb]{\smash{\SetFigFont{12}{14.4}{\rmdefault}{\mddefault}{\updefault}$\ket{1}$}}}
\put(1,-360){\makebox(0,0)[lb]{\smash{\SetFigFont{12}{14.4}{\rmdefault}{\mddefault}{\updefault}$\ket{2}$}}}
\put(1,-810){\makebox(0,0)[lb]{\smash{\SetFigFont{12}{14.4}{\rmdefault}{\mddefault}{\updefault}$\ket{3}$}}}
\put(1,-1260){\makebox(0,0)[lb]{\smash{\SetFigFont{12}{14.4}{\rmdefault}{\mddefault}{\updefault}$\ket{4}$}}}
\put(601, 89){\makebox(0,0)[lb]{\smash{\SetFigFont{12}{14.4}{\rmdefault}{\mddefault}{\updefault}$w_1$}}}
\put(601,-361){\makebox(0,0)[lb]{\smash{\SetFigFont{12}{14.4}{\rmdefault}{\mddefault}{\updefault}$w_2$}}}
\put(601,-811){\makebox(0,0)[lb]{\smash{\SetFigFont{12}{14.4}{\rmdefault}{\mddefault}{\updefault}$w_3$}}}
\put(601,-1261){\makebox(0,0)[lb]{\smash{\SetFigFont{12}{14.4}{\rmdefault}{\mddefault}{\updefault}$w_4$}}}
\put(1201, 89){\makebox(0,0)[lb]{\smash{\SetFigFont{12}{14.4}{\rmdefault}{\mddefault}{\updefault}$w_2$}}}
\put(1201,-361){\makebox(0,0)[lb]{\smash{\SetFigFont{12}{14.4}{\rmdefault}{\mddefault}{\updefault}$w_1$}}}
\put(1201,-811){\makebox(0,0)[lb]{\smash{\SetFigFont{12}{14.4}{\rmdefault}{\mddefault}{\updefault}$w_3$}}}
\put(1201,-1261){\makebox(0,0)[lb]{\smash{\SetFigFont{12}{14.4}{\rmdefault}{\mddefault}{\updefault}$w_4$}}}
\put(1801, 89){\makebox(0,0)[lb]{\smash{\SetFigFont{12}{14.4}{\rmdefault}{\mddefault}{\updefault}$w_2$}}}
\put(1801,-361){\makebox(0,0)[lb]{\smash{\SetFigFont{12}{14.4}{\rmdefault}{\mddefault}{\updefault}$w_3$}}}
\put(1801,-811){\makebox(0,0)[lb]{\smash{\SetFigFont{12}{14.4}{\rmdefault}{\mddefault}{\updefault}$w_1$}}}
\put(1801,-1261){\makebox(0,0)[lb]{\smash{\SetFigFont{12}{14.4}{\rmdefault}{\mddefault}{\updefault}$w_4$}}}
\put(2401, 89){\makebox(0,0)[lb]{\smash{\SetFigFont{12}{14.4}{\rmdefault}{\mddefault}{\updefault}$w_2$}}}
\put(2401,-361){\makebox(0,0)[lb]{\smash{\SetFigFont{12}{14.4}{\rmdefault}{\mddefault}{\updefault}$w_3$}}}
\put(2401,-811){\makebox(0,0)[lb]{\smash{\SetFigFont{12}{14.4}{\rmdefault}{\mddefault}{\updefault}$w_4$}}}
\put(2401,-1261){\makebox(0,0)[lb]{\smash{\SetFigFont{12}{14.4}{\rmdefault}{\mddefault}{\updefault}$w_1$}}}
\put(3001, 89){\makebox(0,0)[lb]{\smash{\SetFigFont{12}{14.4}{\rmdefault}{\mddefault}{\updefault}$w_3$}}}
\put(3001,-361){\makebox(0,0)[lb]{\smash{\SetFigFont{12}{14.4}{\rmdefault}{\mddefault}{\updefault}$w_2$}}}
\put(3001,-811){\makebox(0,0)[lb]{\smash{\SetFigFont{12}{14.4}{\rmdefault}{\mddefault}{\updefault}$w_4$}}}
\put(3001,-1261){\makebox(0,0)[lb]{\smash{\SetFigFont{12}{14.4}{\rmdefault}{\mddefault}{\updefault}$w_1$}}}
\put(3601, 89){\makebox(0,0)[lb]{\smash{\SetFigFont{12}{14.4}{\rmdefault}{\mddefault}{\updefault}$w_3$}}}
\put(3601,-361){\makebox(0,0)[lb]{\smash{\SetFigFont{12}{14.4}{\rmdefault}{\mddefault}{\updefault}$w_4$}}}
\put(3601,-811){\makebox(0,0)[lb]{\smash{\SetFigFont{12}{14.4}{\rmdefault}{\mddefault}{\updefault}$w_2$}}}
\put(3601,-1261){\makebox(0,0)[lb]{\smash{\SetFigFont{12}{14.4}{\rmdefault}{\mddefault}{\updefault}$w_1$}}}
\put(4201, 89){\makebox(0,0)[lb]{\smash{\SetFigFont{12}{14.4}{\rmdefault}{\mddefault}{\updefault}$w_4$}}}
\put(4201,-811){\makebox(0,0)[lb]{\smash{\SetFigFont{12}{14.4}{\rmdefault}{\mddefault}{\updefault}$w_1$}}}
\put(4201,-1261){\makebox(0,0)[lb]{\smash{\SetFigFont{12}{14.4}{\rmdefault}{\mddefault}{\updefault}$w_1$}}}
\put(4201,-361){\makebox(0,0)[lb]{\smash{\SetFigFont{12}{14.4}{\rmdefault}{\mddefault}{\updefault}$w_3$}}}
\put(976,314){\makebox(0,0)[lb]{\smash{\SetFigFont{12}{14.4}{\rmdefault}{\mddefault}{\updefault}$f_1$}}}
\put(3976,314){\makebox(0,0)[lb]{\smash{\SetFigFont{12}{14.4}{\rmdefault}{\mddefault}{\updefault}$f_1$}}}
\put(1576,314){\makebox(0,0)[lb]{\smash{\SetFigFont{12}{14.4}{\rmdefault}{\mddefault}{\updefault}$f_2$}}}
\put(3376,314){\makebox(0,0)[lb]{\smash{\SetFigFont{12}{14.4}{\rmdefault}{\mddefault}{\updefault}$f_2$}}}
\put(2176,314){\makebox(0,0)[lb]{\smash{\SetFigFont{12}{14.4}{\rmdefault}{\mddefault}{\updefault}$f_3$}}}
\put(2776,314){\makebox(0,0)[lb]{\smash{\SetFigFont{12}{14.4}{\rmdefault}{\mddefault}{\updefault}$f_1$}}}
\end{picture}

\end{center}
where $f_m$, $m=1,2,3$, refers to a control pulse of frequency $\omega_m$ with effective 
pulse area $\pi$.  The first pulse interchanges the populations of levels $\ket{1}$ and 
$\ket{2}$, the second pulse flips the populations of levels $\ket{2}$ and $\ket{3}$, the
third pulse switches the populations of levels $\ket{3}$ and $\ket{4}$, etc.  Since the 
populations of our initial ensemble satisfy $w_1<w_2<w_3<w_4$, the energy of the system 
assumes its kinematical minimum at $t=0$ and increases monotonically to its kinematical
maximum value at the final time.  Again, the gradient of approach is more uniform for 
square wave pulses.  

As regards the field strengths, note that for 200 ps pulses up to 5.7 MV/m are required 
for SWP, and up to 12 MV/m for Gaussian pulses, which corresponds to (peak) intensities 
$I=\epsilon_0 c E^2$ of up to $8.5 \mbox{ MW/cm}^2$ (SWP) and $24 \mbox{ MW/cm}^2$ (GWP),
respectively.  Achieving these intensities experimentally should be no problem for pulsed
laser systems.  For CW lasers, it would be challenging at the moment, but it should still
be feasible using a combination of powerful lasers and beam focusing techniques. Moreover,
such problems should disappear with improvements in laser technology in the future.

Had we instead of fixing the pulse length at 200 ps, fixed the strength of the fields to
be $2 A_k = 5 \times 10^6$ V/m, say, then the length $\Delta t_k$ of the control pulses 
according to (\ref{eq:tmax:SWP}) would have been 224.5, 164.6, 138.1, 224.5, 164.6 and 
224.5 ps, respectively, for SWP with $\tau_0=20$ ps.  For GWP with $q_k=4/\Delta t_k$ 
the pulse lengths according to (\ref{eq:tmax:GWP}) would have been 461.3, 326.2, 266.3, 
461.3, 326.3 and 461.3 ps, respectively.  Thus, instead of 1.2 ns in both cases, the time
required to achieve the control objective would have been 1.14 ns for SWP, and 2.3 ns for
GWP.

Note that the problem of population transfer for a system initially in state $\ket{1}$ is
a special case of the problem of population inversion for a trivial ensemble with $w_1=1$
and $w_2=w_3=w_4=0$.  It can easily been seen that pulses four, five and six in the pulse
sequence above do no harm but have no effect for this initial ensemble
\begin{center}
\setlength{\unitlength}{3000sp}%
\begingroup\makeatletter\ifx\SetFigFont\undefined%
\gdef\SetFigFont#1#2#3#4#5{%
  \reset@font\fontsize{#1}{#2pt}%
  \fontfamily{#3}\fontseries{#4}\fontshape{#5}%
  \selectfont}%
\fi\endgroup%
\begin{picture}(3604,1755)(601,-1261)
\thinlines
\put(880,-166){\vector( 1, 1){285}}
\put(880, 52){\vector( 1,-1){285}}
\put(1480,-646){\vector( 1, 1){285}}
\put(1480,-428){\vector( 1,-1){285}}
\put(2080,-1096){\vector( 1, 1){285}}
\put(2080,-878){\vector( 1,-1){285}}
\put(2680,-166){\vector( 1, 1){285}}
\put(2680, 52){\vector( 1,-1){285}}
\put(3280,-661){\vector( 1, 1){285}}
\put(3280,-443){\vector( 1,-1){285}}
\put(3880,-196){\vector( 1, 1){285}}
\put(3880, 22){\vector( 1,-1){285}}
\put(1,90){\makebox(0,0)[lb]{\smash{\SetFigFont{12}{14.4}{\rmdefault}{\mddefault}{\updefault}$\ket{1}$}}}
\put(1,-360){\makebox(0,0)[lb]{\smash{\SetFigFont{12}{14.4}{\rmdefault}{\mddefault}{\updefault}$\ket{2}$}}}
\put(1,-810){\makebox(0,0)[lb]{\smash{\SetFigFont{12}{14.4}{\rmdefault}{\mddefault}{\updefault}$\ket{3}$}}}
\put(1,-1260){\makebox(0,0)[lb]{\smash{\SetFigFont{12}{14.4}{\rmdefault}{\mddefault}{\updefault}$\ket{4}$}}}
\put(601, 89){\makebox(0,0)[lb]{\smash{\SetFigFont{12}{14.4}{\rmdefault}{\mddefault}{\updefault}$1$}}}
\put(601,-361){\makebox(0,0)[lb]{\smash{\SetFigFont{12}{14.4}{\rmdefault}{\mddefault}{\updefault}$0$}}}
\put(601,-811){\makebox(0,0)[lb]{\smash{\SetFigFont{12}{14.4}{\rmdefault}{\mddefault}{\updefault}$0$}}}
\put(601,-1261){\makebox(0,0)[lb]{\smash{\SetFigFont{12}{14.4}{\rmdefault}{\mddefault}{\updefault}$0$}}}
\put(1201, 89){\makebox(0,0)[lb]{\smash{\SetFigFont{12}{14.4}{\rmdefault}{\mddefault}{\updefault}$0$}}}
\put(1201,-361){\makebox(0,0)[lb]{\smash{\SetFigFont{12}{14.4}{\rmdefault}{\mddefault}{\updefault}$1$}}}
\put(1201,-811){\makebox(0,0)[lb]{\smash{\SetFigFont{12}{14.4}{\rmdefault}{\mddefault}{\updefault}$0$}}}
\put(1201,-1261){\makebox(0,0)[lb]{\smash{\SetFigFont{12}{14.4}{\rmdefault}{\mddefault}{\updefault}$0$}}}
\put(1801, 89){\makebox(0,0)[lb]{\smash{\SetFigFont{12}{14.4}{\rmdefault}{\mddefault}{\updefault}$0$}}}
\put(1801,-361){\makebox(0,0)[lb]{\smash{\SetFigFont{12}{14.4}{\rmdefault}{\mddefault}{\updefault}$0$}}}
\put(1801,-811){\makebox(0,0)[lb]{\smash{\SetFigFont{12}{14.4}{\rmdefault}{\mddefault}{\updefault}$1$}}}
\put(1801,-1261){\makebox(0,0)[lb]{\smash{\SetFigFont{12}{14.4}{\rmdefault}{\mddefault}{\updefault}$0$}}}
\put(2401, 89){\makebox(0,0)[lb]{\smash{\SetFigFont{12}{14.4}{\rmdefault}{\mddefault}{\updefault}$0$}}}
\put(2401,-361){\makebox(0,0)[lb]{\smash{\SetFigFont{12}{14.4}{\rmdefault}{\mddefault}{\updefault}$0$}}}
\put(2401,-811){\makebox(0,0)[lb]{\smash{\SetFigFont{12}{14.4}{\rmdefault}{\mddefault}{\updefault}$0$}}}
\put(2401,-1261){\makebox(0,0)[lb]{\smash{\SetFigFont{12}{14.4}{\rmdefault}{\mddefault}{\updefault}$1$}}}
\put(3001, 89){\makebox(0,0)[lb]{\smash{\SetFigFont{12}{14.4}{\rmdefault}{\mddefault}{\updefault}$0$}}}
\put(3001,-361){\makebox(0,0)[lb]{\smash{\SetFigFont{12}{14.4}{\rmdefault}{\mddefault}{\updefault}$0$}}}
\put(3001,-811){\makebox(0,0)[lb]{\smash{\SetFigFont{12}{14.4}{\rmdefault}{\mddefault}{\updefault}$0$}}}
\put(3001,-1261){\makebox(0,0)[lb]{\smash{\SetFigFont{12}{14.4}{\rmdefault}{\mddefault}{\updefault}$1$}}}
\put(3601, 89){\makebox(0,0)[lb]{\smash{\SetFigFont{12}{14.4}{\rmdefault}{\mddefault}{\updefault}$0$}}}
\put(3601,-361){\makebox(0,0)[lb]{\smash{\SetFigFont{12}{14.4}{\rmdefault}{\mddefault}{\updefault}$0$}}}
\put(3601,-811){\makebox(0,0)[lb]{\smash{\SetFigFont{12}{14.4}{\rmdefault}{\mddefault}{\updefault}$0$}}}
\put(3601,-1261){\makebox(0,0)[lb]{\smash{\SetFigFont{12}{14.4}{\rmdefault}{\mddefault}{\updefault}$1$}}}
\put(4201, 89){\makebox(0,0)[lb]{\smash{\SetFigFont{12}{14.4}{\rmdefault}{\mddefault}{\updefault}$0$}}}
\put(4201,-811){\makebox(0,0)[lb]{\smash{\SetFigFont{12}{14.4}{\rmdefault}{\mddefault}{\updefault}$1$}}}
\put(4201,-1261){\makebox(0,0)[lb]{\smash{\SetFigFont{12}{14.4}{\rmdefault}{\mddefault}{\updefault}$1$}}}
\put(4201,-361){\makebox(0,0)[lb]{\smash{\SetFigFont{12}{14.4}{\rmdefault}{\mddefault}{\updefault}$0$}}}
\put(976,314){\makebox(0,0)[lb]{\smash{\SetFigFont{12}{14.4}{\rmdefault}{\mddefault}{\updefault}$f_1$}}}
\put(3976,314){\makebox(0,0)[lb]{\smash{\SetFigFont{12}{14.4}{\rmdefault}{\mddefault}{\updefault}$f_1$}}}
\put(1576,314){\makebox(0,0)[lb]{\smash{\SetFigFont{12}{14.4}{\rmdefault}{\mddefault}{\updefault}$f_2$}}}
\put(3376,314){\makebox(0,0)[lb]{\smash{\SetFigFont{12}{14.4}{\rmdefault}{\mddefault}{\updefault}$f_2$}}}
\put(2176,314){\makebox(0,0)[lb]{\smash{\SetFigFont{12}{14.4}{\rmdefault}{\mddefault}{\updefault}$f_3$}}}
\put(2776,314){\makebox(0,0)[lb]{\smash{\SetFigFont{12}{14.4}{\rmdefault}{\mddefault}{\updefault}$f_1$}}}
\end{picture}

\end{center}
and can therefore be omitted.  Thus, the general six pulse sequence simplifies in this case
to the three pulse sequence in the previous section.  This can also be inferred directly 
from the decomposition (\ref{eq:Udecomp2}) of the target operator.  For a four level system
(\ref{eq:Udecomp2}) becomes $\op{U}=\op{U}_0(T)\op{V}_1\op{V}_2\op{V}_1\op{V}_3\op{V}_2
\op{V}_1$ with $\op{V}_m$ as in (\ref{eq:Vm}).  Thus, after applying the pulse sequence the
state of the system is
\begin{eqnarray*}
  \op{\rho} &=& \op{U}_0(T)\op{V}_1\op{V}_2\op{V}_1\op{V}_3\op{V}_2\op{V}_1 \op{\rho}_0
                [\op{U}_0(T)\op{V}_1\op{V}_2\op{V}_1\op{V}_3\op{V}_2\op{V}_1]^\dagger \\
            &=& \op{U}_0(T)\op{V}_1\op{V}_2\op{V}_1\op{V}_3\op{V}_2\op{V}_1 \op{\rho}_0
                \op{V}_1^\dagger \op{V}_2^\dagger\op{V}_3^\dagger
                \op{V}_1^\dagger\op{V}_2^\dagger\op{V}_1^\dagger\op{U}_0(T)^\dagger.
\end{eqnarray*}
If $\op{\rho}_0=\ket{1}\bra{1}$ then $\op{V}_3\op{V}_2\op{V}_1 \op{\rho}_0\op{V}_1^\dagger
\op{V}_2^\dagger\op{V}_3^\dagger=\ket{4}\bra{4}$.  Since $\op{U}_0(T)\op{V}_1\op{V}_2
\op{V}_1$ commutes with this operator, the remaining factors cancel in the decomposition
\begin{eqnarray*}
 & & \op{U}_0(T)\op{V}_1\op{V}_2\op{V}_1\op{V}_3\op{V}_2\op{V}_1 \op{\rho}_0
                \op{V}_1^\dagger\op{V}_2^\dagger\op{V}_3^\dagger
                \op{V}_1^\dagger\op{V}_2^\dagger\op{V}_1^\dagger\op{U}_0(T)^\dagger\\
 &=& \op{U}_0(T)\op{V}_1\op{V}_2\op{V}_1\ket{4}\bra{4}
                \op{V}_1^\dagger\op{V}_2^\dagger\op{V}_1^\dagger\op{U}_0(T)^\dagger\\
 &=& \ket{4}\bra{4}\op{U}_0(T)\op{V}_1\op{V}_2\op{V}_1\op{V}_1^\dagger\op{V}_2^\dagger
     \op{V}_1^\dagger\op{U}_0(T)^\dagger = \ket{4}\bra{4}
\end{eqnarray*}
and hence do not change the state of the system.

\section{Creation of arbitrary superposition states}
\label{sec:superposition}

In this section we consider the problem of creating arbitrary superposition states from
an initial energy eigenstate.  Control schemes to create such superposition states may be
useful in controlling quantum interference in multi-state systems, and can be considered
a generalization of the $\frac{\pi}{2}$ pulses used routinely in free induction-decay 
experiments \cite{61Abraham}.

Concretely, assume that the system is initially in the ground state $\ket{1}$.  To create
the superposition state  
\begin{equation}\label{eq:superpos}
  \ket{\Psi(t)} = \sum_{n=1}^N r_n e^{\rmi\theta_n} e^{\rmi E_n t/\hbar}\ket{n}
                = \sum_{n=1}^N r_n e^{\rmi\theta_n} \ket{\tilde{n}(t)}
\end{equation}
where the coefficients $r_n$ satisfy the normalization condition $\sum_{n=1}^N r_n^2=1$,
we need to find a unitary operator $\op{U}_I$ such that 
\begin{equation} \label{eq:UI}
  \op{U}_I \ket{1} = \sum_{n=1}^N r_n e^{\rmi\theta_n} \ket{n}
\end{equation}
and decompose $\op{U}_I$ according to the algorithm described in \ref{appendix:Udecomp}.

To find a unitary operator $\op{U}_I$ that satisfies (\ref{eq:U1}) we set
\begin{equation}
  \op{W} =\left( \begin{array}{c|c} 
                r_1 & \vec{0} \\\hline
                r_2 &  \\
             \vdots &  \op{I}_{N-1}\\
                r_N & \\ 
               \end{array} \right),
\end{equation}
where $\op{I}_{N-1}$ is the identity matrix in dimension $N-1$, and perform Gram-Schmidt
orthonormalization on the columns of $\op{W}$.  This produces a matrix $\op{U}_1$ which
is unitary and satisfies $\op{U}_1\ket{1}=\sum_{n=1}^N r_n \ket{n}$.  Hence, $\op{U}_I=
\op{\Theta}\op{U}_1$ with $\op{\Theta}=\sum_{n=1}^N e^{\rmi\theta_n}\ket{n}\bra{n}$ 
satisfies (\ref{eq:UI}).  

As an example, we consider the problem of creating the superposition state $\ket{\Psi(t)}
=\frac{1}{2}\sum_{n=1}^4 \ket{\tilde{n}(t)}$ for a four-level system initially in state 
$\ket{1}$.  As outlined above, we set
\begin{equation}
  \op{W} =\left(\begin{array}{cccc} 
                1/2 & 0 & 0 & 0 \\
                1/2 & 1 & 0 & 0 \\
                1/2 & 0 & 1 & 0 \\
                1/2 & 0 & 0 & 1
                \end{array} \right).
\end{equation}
and perform Gram-Schmidt orthonormalization on the columns of $\op{W}$, which gives
\begin{equation} \label{eq:targetU1}
  \op{U}_1=\left(\begin{array}{cccc}
   1/2 & -\sqrt{3}/6 & -\sqrt{6}/6 & -\sqrt{2}/2 \\
   1/2 & +\sqrt{3}/2 & 0           & 0 \\
   1/2 & -\sqrt{3}/6 & +\sqrt{6}/3 & 0 \\
   1/2 & -\sqrt{3}/6 & -\sqrt{6}/6 & +\sqrt{2}/2
  \end{array}\right).
\end{equation}
Since $\op{\Theta}=\op{I}$ we have $\op{U}_I = \op{U}_1$ and applying the decomposition 
algorithm (appendix \ref{appendix:Udecomp}) leads to the factorization $\op{U}_1 = 
\op{V}_5 \op{V}_4 \op{V}_3 \op{V}_2 \op{V}_1$, where the factors are
\begin{equation} \label{eq:Udecomp3b}
\begin{array}{rcl}
  \op{V}_1 &=& \exp\left(+C_1\op{x}_1 \right), \quad C_1 = \frac{\pi}{3},\\
  \op{V}_2 &=& \exp\left(-C_2\op{x}_2 \right), \quad C_2 = \arctan\left(\sqrt{2}\right) \\
  \op{V}_3 &=& \exp\left(+C_3\op{x}_3 \right), \quad C_3 = \frac{\pi}{4},\\
  \op{V}_4 &=& \exp\left(+C_4\op{x}_2 \right), \quad C_4 = \frac{\pi}{2},\\
  \op{V}_5 &=& \exp\left(-C_5\op{x}_1 \right), \quad C_5 = \frac{\pi}{2}.
\end{array}
\end{equation}
This decomposition corresponds to the following sequence of five control pulses
\[\begin{array}{rll}
  f_1(t) &= A_1(t) e^{i(\omega_1 t +\pi/2)} + \mbox{c.c.} &= -2A_1(t) \sin(\omega_1 t) \\
  f_2(t) &= A_2(t) e^{i(\omega_2 t -\pi/2)} + \mbox{c.c.} &= +2A_2(t) \sin(\omega_2 t) \\ 
  f_3(t) &= A_3(t) e^{i(\omega_3 t +\pi/2)} + \mbox{c.c.} &= -2A_3(t) \sin(\omega_3 t) \\ 
  f_4(t) &= A_4(t) e^{i(\omega_2 t +\pi/2)} + \mbox{c.c.} &= -2A_4(t) \sin(\omega_2 t) \\ 
  f_5(t) &= A_5(t) e^{i(\omega_1 t -\pi/2)} + \mbox{c.c.} &= +2A_5(t) \sin(\omega_1 t) \\ 
\end{array}\]
with pulse areas $\frac{2}{3}\pi$, $2\arctan(\sqrt{2})$, $\frac{1}{2}\pi$, $\pi$ and $\pi$,
respectively.  Note that only five instead of six pulses are required since the target 
operator $\op{U}_1$ has two consecutive zeros in the last column, which implies that one
of the six control pulses has zero amplitude and can thus be omitted.  

\begin{figure}
\begin{center}
\begin{tabular}{l@{}l}
(a) Square wave pulses & (b) Gaussian pulses \\
\myincludegraphics[width=3in]{figures/pdf/Rb4_Psi1_SWP.pdf}{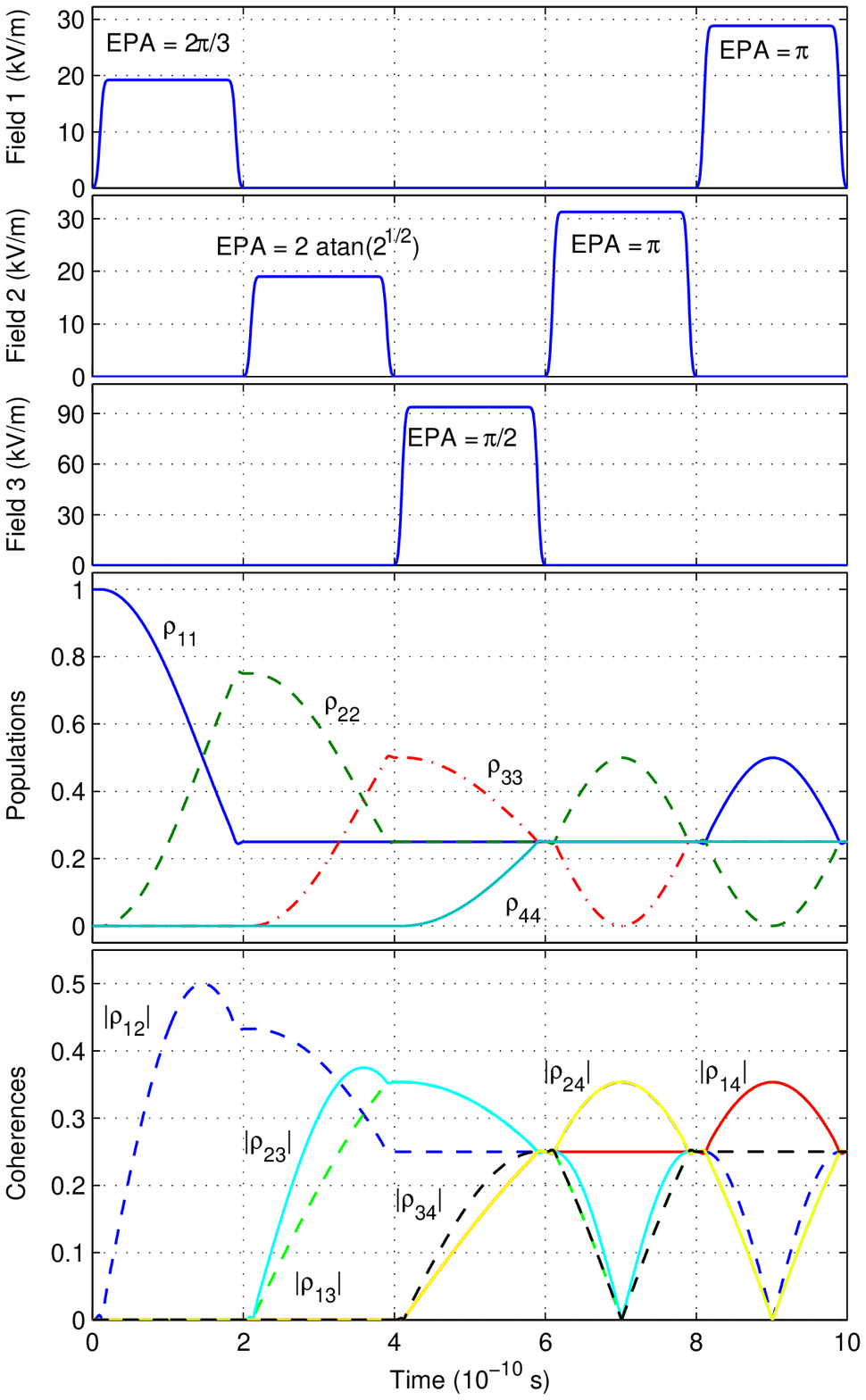} 
&
\myincludegraphics[width=3in]{figures/pdf/Rb4_Psi1_GWP.pdf}{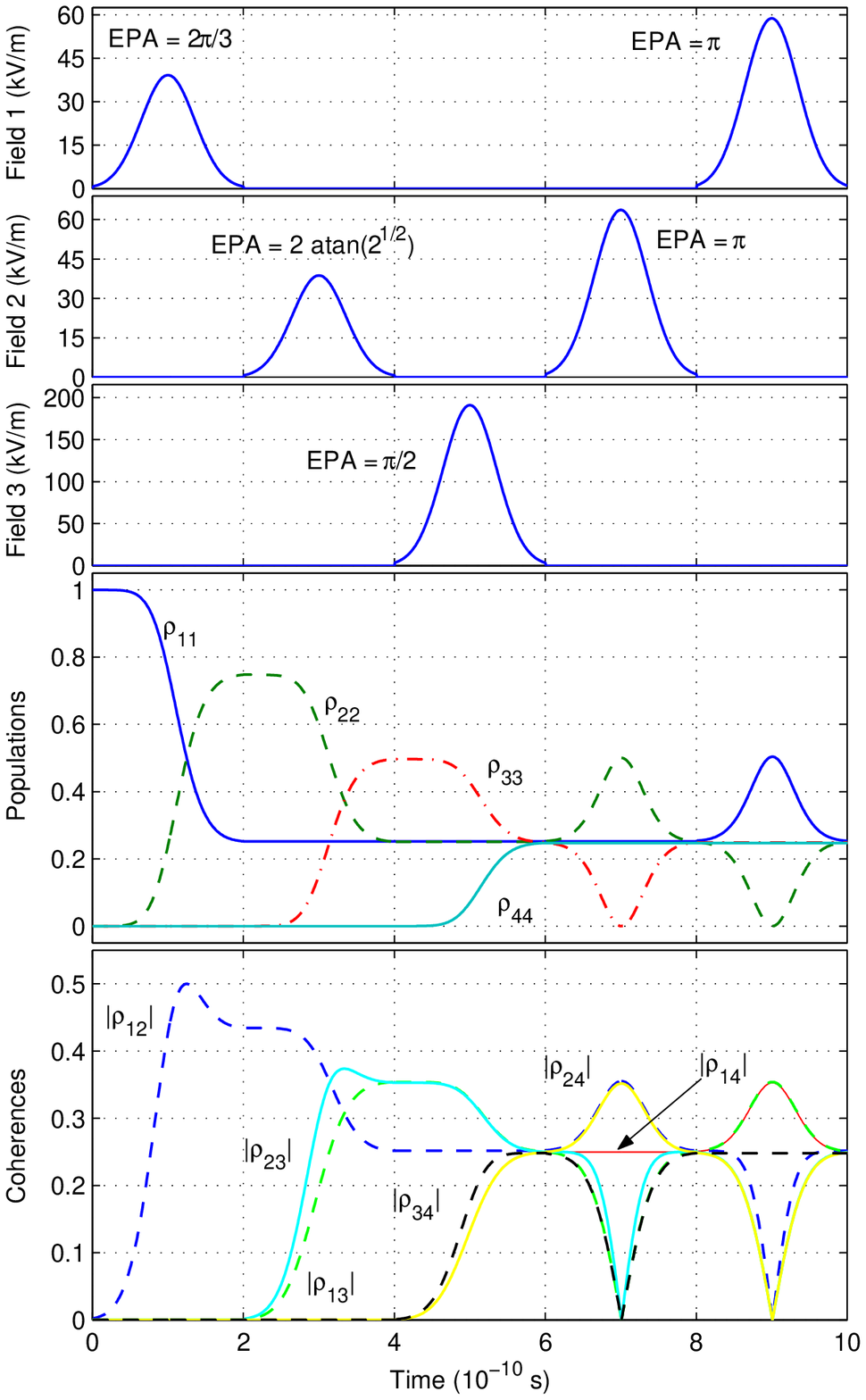}
\end{tabular}
\end{center}
\caption{Creation of the superposition state $\ket{\Psi(t)}=\frac{1}{2}\sum_{n=1}^4
\ket{\tilde{n}(t)}$ for \Rb87 initially in the ground state $\ket{1}$ using (a) five 
200 ps square wave pulses with rise and decay time $\tau_0=20$ ps, and (b) five 200 ps 
Gaussian pulses with $q=2\times 10^{10}$ Hz.  The top graphs show the pulse envelopes 
$A_k(t)$.  The effective pulse area (EPA) of all pulses is as shown in the graph.  The 
labels `Field $m$' indicate that the corresponding pulses are resonant with the frequency
$\omega_m$ of the transition $\ket{m}\rightarrow\ket{m+1}$.}
\label{Fig:Superpos} 
\end{figure}

Figure \ref{Fig:Superpos} shows the results of a control simulation based on this
decomposition of $\op{U}_1$ for \Rb87 using square wave and Gaussian control pulses, 
respectively.  Note that all the populations and the absolute values of all the 
coherences are $0.25$ at the final time --- exactly as required for the superposition
state $\ket{\Psi(t)}=\frac{1}{2}\sum_{n=1}^4 \ket{\tilde{n}(t)}$, whose density matrix
representation is 
\[
 \op{\rho}(t) =\ket{\Psi(t)}\bra{\Psi(t)} 
              = \frac{1}{4} 
                \left( \begin{array}{cccc} 
                1 & e^{\rmi\omega_{12}t} & e^{\rmi\omega_{13}t} & e^{\rmi\omega_{14}t} \\
                e^{-\rmi\omega_{12}t} & 1 & e^{\rmi\omega_{23}t}& e^{\rmi\omega_{24}t} \\
                e^{-\rmi\omega_{13}t} & e^{-\rmi\omega_{23}t} & 1 & e^{\rmi\omega_{34}t}\\
                e^{-\rmi\omega_{14}t} & e^{-\rmi\omega_{24}t} & e^{-\rmi\omega_{34}t} &1 
                 \end{array}\right),
\]
i.e., $|\op{\rho}_{mn}|=\frac{1}{4}$ for all $m,n$.  Note that we have plotted the absolute
values of the coherences $\op{\rho}_{mn}(t)$ (for $m\neq n$) only since their phases are 
rapidly oscillating at frequencies $\omega_{mn}=(E_n-E_m)/\hbar$, which are on the order
of $10^{15}$ Hz for \Rb87.  The pulse intensities are similar to those for population 
transfer in \Rb87.  Again, we chose pulses of fixed length 200 ps.  Had we instead fixed 
the strength of the fields to be $2 A_k = 10^5$ V/m, say, then the length $\Delta t_k$ of 
the control pulses according to (\ref{eq:tmax:SWP}) and (\ref{eq:tmax:GWP}) would have been
99.5, 88.6, 358.6, 132.9 and 155.4 ps, respectively, for SWP with $\tau_0=20$ ps, and 156.7, 
154.9, 764.1, 254.7 and 305.6 ps, respectively, for GWP with $q_k=4/\Delta t_k$.  

Unlike decompositions (\ref{eq:Udecomp1}) and (\ref{eq:Udecomp2}) where the initial phases
$\phi_m$ of the control pulses were arbitrary, the factorization (\ref{eq:Udecomp3b}) fixes
the pulse area and frequency $\omega_m$ as well as the initial phase $\phi_m$ of each pulse.
In order to determine the significance of the the initial pulse phases on the outcome of 
the control process, we compute the unitary operator $\op{U}_2=\tilde{V}_5 \tilde{V}_4 
\tilde{V}_3 \tilde{V}_2 \tilde{V}_1$, where the factors are
\begin{equation} \label{eq:Udecomp3c}
\begin{array}{rcl}
  \tilde{V}_1 &=& \exp\left[C_1(\sin\phi_1\op{x}_1-\cos\phi_1\op{y}_1)\right] \\
  \tilde{V}_2 &=& \exp\left[C_2(\sin\phi_2\op{x}_2-\cos\phi_2\op{y}_2)\right] \\
  \tilde{V}_3 &=& \exp\left[C_3(\sin\phi_3\op{x}_3-\cos\phi_3\op{y}_3)\right] \\
  \tilde{V}_4 &=& \exp\left[C_4(\sin\phi_4\op{x}_2-\cos\phi_4\op{y}_2)\right] \\
  \tilde{V}_5 &=& \exp\left[C_5(\sin\phi_5\op{x}_1-\cos\phi_5\op{y}_1)\right]
\end{array}
\end{equation}
and the constants $C_k$ are as in (\ref{eq:Udecomp3b}) but the initial phases $\phi_k$ of
the control pulses are arbitrary, and apply this operator to the initial state $\ket{1}$.
The resulting state 
\begin{equation} \label{eq:phi}
  \op{U}_2 \left(\begin{array}{c} 1 \\ 0 \\ 0 \\ 0 \end{array} \right)
= \frac{1}{2}
  \left(\begin{array}{l} e^{\rmi(\phi_4+\phi_5-\phi_1-\phi_2)} \\ 
                         e^{\rmi(-\pi/2-\phi_5) } \\ 
                         e^{\rmi(\pi-\phi_1-\phi_4)} \\ 
                         e^{\rmi(\pi/2-\phi_1-\phi_2-\phi_3)} 
         \end{array} \right)
\end{equation}
differs from the desired target state only in the phase factors, i.e., the pulse phases 
do not affect the relative amplitudes $r_n$ of the superposition state created.  Moreover,
we can use (\ref{eq:phi}) to explicitly determine the pulse phases $\phi_n$ as a function 
of the phases $\theta_n$ of the target state: 
\begin{equation}
\begin{array}{rcl}
  \phi_1 &=& \mbox{arbitrary} \\
  \phi_2 &=& \frac{\pi}{2} - 2\phi_1 - \theta_1 - \theta_2 - \theta_3 \\
  \phi_3 &=& \phi_1 + \theta_1 + \theta_2 + \theta_3 - \theta_4 \\
  \phi_4 &=& \pi - \phi_1 - \theta_3 \\
  \phi_5 &=& -\frac{\pi}{2} - \theta_2. 
\end{array}
\end{equation} 
Setting $\theta_n=0$ for $n=1,2,3,4$ and choosing $\phi_1=\pi/2$ leads to $\phi_2=-\pi/2$,
$\phi_3=\pi/2$, $\phi_4=\pi/2$ and $\phi_5=-\pi/2$, which agrees with the phases in 
decomposition (\ref{eq:Udecomp3b}).  

\section{Optimization of observables}
\label{sec:optimization}

Finally, we address the problem of maximizing the ensemble average of an observable for a
system whose initial state is a statistical ensemble of energy eigenstates (\ref{eq:rho0}).
Let us first consider the case of a time-independent observable $\op{A}$.  To determine 
the target operator required to maximize the ensemble average $\ave{\op{A}}$ of $\op{A}$ 
we observe that $\ave{\op{A}}$ is bounded above by the kinematical upper bound 
\cite{PRA58p2684}
\begin{equation}
  \ave{\op{A}} \le \sum_{n=1}^N w_{\sigma(n)} \lambda_n, 
\end{equation}
where $\lambda_n$ are the eigenvalues of $\op{A}$ counted with multiplicity and ordered
in a non-increasing sequence
\begin{equation}
   \lambda_1 \ge \lambda_2 \ge \cdots \ge \lambda_N, 
\end{equation}
$w_n$ are the populations of the energy levels $E_n$ of the initial ensemble, and $\sigma$
is a permutation of $\{1,\ldots,N\}$ such that
\begin{equation}
   w_{\sigma(1)} \ge w_{\sigma(2)} \ge \cdots \ge w_{\sigma(N)}.
\end{equation}
Observe that this universal upper bound for the ensemble average of any observable 
$\op{A}$ is dynamically attainable since the systems considered in this paper are 
completely controllable \cite{PRA63n025403, PRA63n063410}.  

Let $\ket{\Psi_n}$ for $1\le n\le N$ denote the normalized eigenstates of $\op{A}$ 
satisfying $\op{A}\ket{\Psi_n}=\lambda_n \ket{\Psi_n}$ and let $\op{U}_1$ be a unitary 
transformation such that
\begin{equation} \label{eq:U1def}
    \ket{\Psi_{\sigma(n)}}=\op{U}_1\ket{n}, \quad 1\le n \le N.
\end{equation}
Given an initial state $\op{\rho}_0$ of the form (\ref{eq:rho0}), we have
\begin{eqnarray}
  \Tr \left(\op{A}\op{U}_1\rho_0 \op{U}_1^\dagger\right) 
  &=& \Tr\left(\op{A}\sum_n w_n \op{U}_1\ket{n}\bra{n}\op{U}_1^\dagger\right) 
  \nonumber \\
  &=& \Tr\left(\sum_n w_n \op{A}\ket{\Psi_{\sigma(n)}}\bra{\Psi_{\sigma(n)}}\right)
  \nonumber\\
  &=& \Tr\left(\sum_n w_n\lambda_{\sigma(n)}\ket{\Psi_{\sigma(n)}}
                                            \bra{\Psi_{\sigma(n)}}\right)
  \nonumber\\
  &=& \sum_n w_n \lambda_{\sigma(n)} = \sum_n w_{\sigma(n)} \lambda_n.
\end{eqnarray}
Hence, if the system is initially in the state (\ref{eq:rho0}) then $\op{U}_1$ is a target
operator for which the observable $\op{A}$ assumes its kinematical maximum, and we can use
the decomposition algorithm described in \ref{appendix:Udecomp} to obtain the required 
factorization of the operator $\op{U}_I=\op{U}_0(T)^\dagger\op{U}_1$.

However, if $\op{A}$ is an observable whose eigenstates are not energy eigenstates then
the expectation value or ensemble average of $\op{A}$ will usually oscillate rapidly as 
a result of the action of the free evolution operator $\op{U}_0(t)$.  These oscillations
are rarely significant for the application at hand and often distracting.  In such cases 
it is advantageous to define a dynamic observable
\begin{equation} 
  \tilde{A}(t) = \op{U}_0(t) \op{A} \op{U}_0(t)^\dagger
\end{equation}
and optimize its ensemble average instead. To accomplish this, note that if $\ket{\Psi_n}$
are the eigenstates of $\op{A}$ satisfying $\op{A}\ket{\Psi_n}=\lambda_n\ket{\Psi_n}$ then
$\ket{\tilde{\Psi}_n(t)}=\op{U}_0(t)\ket{\Psi_n}$ are the corresponding eigenstates of 
$\tilde{A}(t)$ since
\[
   \tilde{A}(t) \ket{\tilde{\Psi}_n(t)}
 = \op{U}_0(t)\op{A} \op{U}_0(t)^\dagger \op{U}_0(t) \ket{\Psi_n}
 = \op{U}_0(t)\lambda_n \ket{\Psi_n} 
 = \lambda_n \ket{\tilde{\Psi}_n(t)}.
\]
Hence, if $\op{U}_1$ is a unitary operator such that equation (\ref{eq:U1def}) holds then
$\op{U}_0(t)\op{U}_1$ is a unitary operator that maps the energy eigenstates $\ket{n}$
onto the $\tilde{A}(t)$-eigenstates $\ket{\tilde{\Psi}_n(t)}$ since
\[
 \op{U}_0(t) \op{U}_1 \ket{n}
 = \op{U}_0(t) \ket{\Psi_{\sigma(n)}}
 = \ket{\tilde{\Psi}_{\sigma(n)}(t)} 
\]
for $1\le n\le N$.  Thus, the evolution operator required to maximize the ensemble average
of $\tilde{A}(t)$ at time $T>0$ is $\op{U}_0(T)\op{U}_1$ and the target operator to be 
decomposed is $\op{U}=\op{U}_0(T)^\dagger\op{U}_0(T)\op{U}_1=\op{U}_1$.

For instance, suppose we wish to maximize the ensemble average of the transition dipole 
moment operator $\tilde{A}(t)=\op{U}_0(t)\op{A}\op{U}_0(t)^\dagger$, where
\begin{equation}\label{eq:A}
  \op{A}= \sum_{n=1}^{N-1} d_n \left(\ket{n}\bra{n+1}+\ket{n+1}\bra{n}\right),
\end{equation}
for a system initially in state (\ref{eq:rho0}) with 
\begin{equation}
  w_1 > w_2 > \cdots > w_N > 0.
\end{equation}
First, we need to find a unitary operator that maps the initial state $\ket{n}$ onto the 
$\op{A}$-eigenstate $\ket{\Psi_n}$ for $1\le n\le N$.  Let $\op{U}_1$ be the $N\times N$
matrix whose $n$th column is the normalized $\op{A}$-eigenstate $\ket{\Psi_n}$.  Then 
$\op{U}_1$ clearly satisfies $\op{U}_1\ket{n}=\ket{\Psi_n}$.  Furthermore, $\op{U}_1$ is
automatically unitary since the eigenstates $\ket{\Psi_n}$ are orthonormal by hypothesis.

For $N=4$ and $d_n=p_0\sqrt{n}$ the eigenvalues of the operator $\op{A}$ defined in 
(\ref{eq:A}) are (in decreasing order) 
\[
 \lambda_1=\sqrt{3+\sqrt{6}}, \;
 \lambda_2=\sqrt{3-\sqrt{6}}, \;
 \lambda_3=-\lambda_2, \;
 \lambda_4=-\lambda_1
\] 
and the corresponding eigenstates with respect to the standard basis $\ket{n}$ are the
columns of the operator 
\begin{equation}
 \op{U}_1=\left[ \begin{array}{cccc} 
 \frac{1}{2\lambda_1}&\frac{1}{2\lambda_2}&\frac{1}{2\lambda_2}&\frac{1}{2\lambda_1}\\[1.ex]
\frac{1}{2}         &\frac{1}{2}         &-\frac{1}{2}        &-\frac{1}{2} \\[1.ex]
 \frac{\sqrt{2}+\sqrt{3}}{2\lambda_1} & \frac{\sqrt{2}-\sqrt{3}}{2\lambda_2} &
 \frac{\sqrt{2}-\sqrt{3}}{2\lambda_2} & \frac{\sqrt{2}+\sqrt{3}}{2\lambda_1} \\[1.ex]
 \frac{1}{2}         &-\frac{1}{2}         &\frac{1}{2}        &-\frac{1}{2} \\
\end{array} \right].
\end{equation}

Applying the decomposition algorithm described in \ref{appendix:Udecomp} yields the 
product decomposition $\op{U}_1 \op{\Theta}=\op{V}_6 \op{V}_5 \op{V}_4 \op{V}_3 \op{V}_2 
\op{V}_1$, where the factors are
\begin{equation} \label{eq:Cs}
   \begin{array}{ll}
  \op{V}_1 = \exp\left(-C_1\op{x}_1 \right), & C_1 = \pi/4,\\
  \op{V}_2 = \exp\left(-C_2\op{x}_2 \right), & C_2 = \arctan\left(\sqrt{2}\right),\\
  \op{V}_3 = \exp\left(-C_3\op{x}_1 \right), & C_3 = 
                       \mbox{arccot}\left(\frac{\sqrt{6}-\sqrt{3}+3\sqrt{2}}{3}\right),\\
  \op{V}_4 = \exp\left(-C_4\op{x}_3 \right), & C_4 = \pi/3,\\
  \op{V}_5 = \exp\left(-C_5\op{x}_2 \right), & C_5 = 
                       \arctan\left(\frac{\sqrt{4+\sqrt{6}}}{\sqrt{2}+\sqrt{3}}\right),\\
  \op{V}_6 = \exp\left(-C_6\op{x}_1 \right), & C_5 = 
                        \mbox{arccot}\left(\sqrt{3+\sqrt{6}}\right)
\end{array}
\end{equation}
and $\op{\Theta}=\mbox{diag}(1,-1,1,-1)$.  Note that $\op{U}_2\equiv\op{U}_1\op{\Theta}$
is equivalent to $\op{U}_1$ since $\op{\Theta}$ commutes with $\op{\rho}_0$ as defined in
equation (\ref{eq:rho0}), i.e., $\op{\Theta}\op{\rho}_0\op{\Theta}^\dagger=\op{\rho}_0$,
and thus
\begin{equation}
 \Tr\left(\op{A}\op{U}_2\op{\rho}_0\op{U}_2^\dagger\right) 
 = \Tr\left(\op{A}\op{U}_1\op{\Theta}\op{\rho}_0\op{\Theta}^\dagger\op{U}_1^\dagger\right)
 =\Tr\left(\op{A}\op{U}_1 \op{\rho}_0\op{U}_1\right). \label{eq:Theta-equiv}
\end{equation}
This decomposition corresponds to a sequence of six control pulses
\[\begin{array}{rll}
  f_1(t) &= A_1(t) e^{\rmi(\omega_1 t -\pi/2)} + \mbox{c.c.} &= 2A_1(t) \sin(\omega_1 t) \\
  f_2(t) &= A_2(t) e^{\rmi(\omega_2 t -\pi/2)} + \mbox{c.c.} &= 2A_2(t) \sin(\omega_2 t) \\ 
  f_3(t) &= A_3(t) e^{\rmi(\omega_1 t -\pi/2)} + \mbox{c.c.} &= 2A_3(t) \sin(\omega_1 t) \\ 
  f_4(t) &= A_4(t) e^{\rmi(\omega_3 t -\pi/2)} + \mbox{c.c.} &= 2A_4(t) \sin(\omega_3 t) \\ 
  f_5(t) &= A_5(t) e^{\rmi(\omega_2 t -\pi/2)} + \mbox{c.c.} &= 2A_5(t) \sin(\omega_2 t) \\ 
  f_6(t) &= A_6(t) e^{\rmi(\omega_1 t -\pi/2)} + \mbox{c.c.} &= 2A_6(t) \sin(\omega_1 t)  
\end{array}\]
with effective pulse areas $\frac{\pi}{2}$, $2C_2$, $2C_3$, $\frac{2\pi}{3}$, $2C_5$ and 
$2C_6$, respectively.  Again, the decomposition fixes the frequency and pulse area as well
as the initial phase of each pulse and the question thus arises what role the phases play.
As we have already seen, the target operator $\op{U}_1$ is not unique.  In fact, equation 
(\ref{eq:Theta-equiv}) shows that right multiplication of $\op{U}_1$ by any unitary matrix
that commutes with $\op{\rho}_0$ produces another unitary operator that leads to the same
ensemble average of the target observable.  Nevertheless, in general, the control process
is sensitive to the phases $\phi_m$.  For instance, one can verify that changing the phase
$\phi_1$ of the first pulse from $-\pi/2$ to $\pi/2$ in the pulse sequence above leads to
the following evolution operator
\begin{equation}
 \op{U}_3 = \left[ \begin{array}{cccc} 
 \frac{1}{2\lambda_2}&\frac{1}{2\lambda_1}&\frac{1}{2\lambda_2}&-\frac{1}{2\lambda_1}\\[1ex]
 \frac{1}{2}         &\frac{1}{2}         &-\frac{1}{2}        &\frac{1}{2} \\[1ex]
 \frac{\sqrt{2}-\sqrt{3}}{2\lambda_2} & \frac{\sqrt{2}+\sqrt{3}}{2\lambda_1} & 
 \frac{\sqrt{2}-\sqrt{3}}{2\lambda_2} & \frac{\sqrt{2}+\sqrt{3}}{-2\lambda_1} \\[1ex]
 -\frac{1}{2}        &\frac{1}{2}     & \frac{1}{2}            & \frac{1}{2} 
\end{array} \right],
\end{equation}
which maps $\ket{3}$ onto $\ket{\Psi_3}$ and $\ket{4}$ onto $-\ket{\Psi_4}$ but $\ket{1}$ 
onto $\ket{\Psi_2}$ and $\ket{2}$ onto $\ket{\Psi_1}$ and leads to the ensemble average
\begin{equation}
 \ave{\op{A}} = w_1 \lambda_2 + w_2 \lambda_1 + w_3 \lambda_3 + w_4 \lambda_4
\end{equation}
at the final time, which is strictly less than the kinematical maximum if $w_1>w_2$.

\begin{figure}
\begin{center}
\begin{tabular}{l@{}l}
(a) Square wave pulses & (b) Gaussian pulses \\
\myincludegraphics[width=3in]{figures/pdf/HF4_Dipole_SWP.pdf}{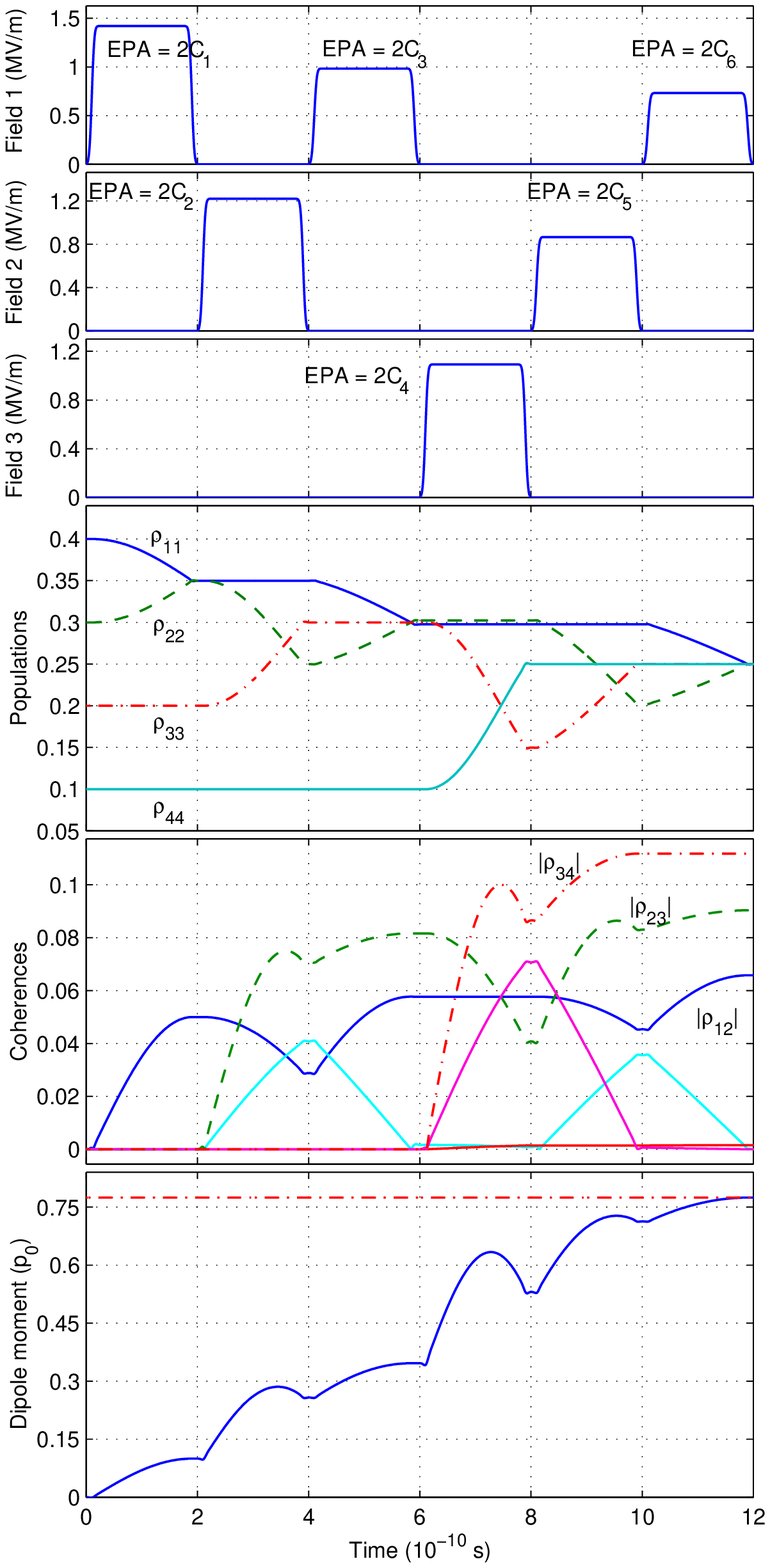}
&
\myincludegraphics[width=3in]{figures/pdf/HF4_Dipole_GWP.pdf}{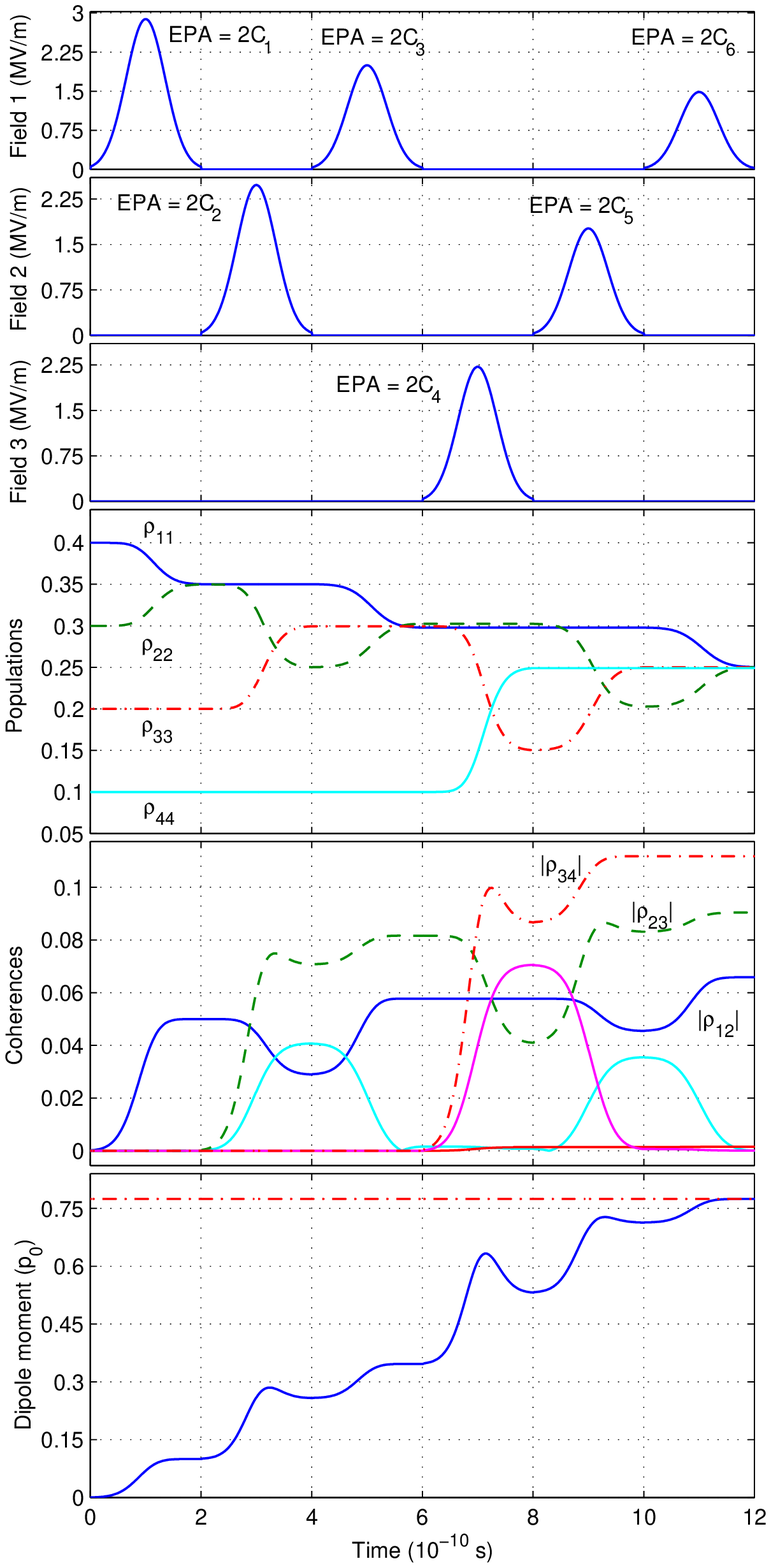}
\end{tabular}
\end{center}
\caption{Maximization of the transition dipole moment for HF using (a) six square wave 
pulses with rise and decay time $\tau_0=20$ ps, and (b) six Gaussian pulses with $q=2
\times 10^{10}$ Hz (right).  The top graphs show the pulse envelopes $A_k(t)$.  The values
of the constants $C_k$ which determine the effective pulse areas (EPA) are given in 
(\ref{eq:Cs}).  The labels `Field $m$' indicate that the corresponding pulses are resonant
with the frequency $\omega_m$ of the transition $\ket{m} \rightarrow \ket{m+1}$.} 
\label{Fig:Dipole} 
\end{figure}

Figure \ref{Fig:Dipole} shows the results of control simulations for HF with initial 
populations $w_1=0.4$, $w_2=0.3$, $w_3=0.2$ and $w_4=0.1$ for square wave and Gaussian 
control pulses, respectively.  The pulse intensities are similar to those for population 
inversion in HF.  Notice that the observable indeed attains its kinematical upper bound
at the final time, as desired.  Furthermore, the target state for which the observable 
assumes its upper bound is
\[
  \op{\rho} = \op{U}_1 \op{\rho}_0 \op{U}_1^\dagger
            =  \left( \begin{array}{cccc}
                 \frac{1}{4} & \rho_{12} & 0 & \rho_{14} \\ 
                 \rho_{12}^\dagger & \frac{1}{4} & \rho_{23} & 0 \\
                 0 & \rho_{23}^\dagger & \frac{1}{4} & \rho_{34} \\
                 \rho_{14}^\dagger & 0 & \rho_{34}^\dagger & \frac{1}{4}  
                \end{array} \right)
\]
with $\rho_{12}=\lambda_1\lambda_2(\lambda_2+\lambda_1/3)/40 \approx 0.0658$,
$\rho_{14}=\lambda_1\lambda_2(\lambda_2-\lambda_1/3)/40 \approx -0.0016$,
$\rho_{23}=\lambda_2(\lambda_1^2-1/\sqrt{3})/40 \approx 0.0904$,
$\rho_{34}=\lambda_2(\lambda_1^2+1/\sqrt{3})/40 \approx 0.1118$, which agrees with the
final values of the populations and coherences in figure \ref{Fig:Dipole}.  Note that we
chose pulses of fixed length 200 ps.  Had we instead fixed the strength of the fields to
be $2 A_k = 5 \times 10^6$ V/m, say, then the length $\Delta t_k$ of the control pulses 
according to (\ref{eq:tmax:SWP}) and (\ref{eq:tmax:GWP}) would have been 122.2, 97.9, 
60.8, 156.3, 82.5 and 72.7 ps, respectively, for SWP with $\tau_0=20$ ps, and 230.6, 
198.4, 92.2, 307.5, 141.0 and 118.9 ps, respectively, for GWP with $q_k=4/\Delta t_k$.

\section{Conclusion}
\label{sec:conclusion}

We have presented several control schemes designed to achieve control objectives ranging 
from population transfer and inversion of ensemble populations to the creation of arbitrary
superposition states and the optimization of (dynamic) observables.  A key feature of these
schemes is that they rely only on sequences of simple control pulses such as square wave 
pulses with finite rise and decay times or Gaussian wavepackets to achieve the control
objective.  In the optical regime, for instance, such pulses can easily be created in the 
laboratory using pulsed laser sources, or by modulating the amplitude of CW lasers using 
Pockel cells.  No sophisticated pulse shaping technology is required.  A limitation of the
approach is the need to be able to selectively address individual transitions, which 
restricts the application of this technique to systems where selection rules and frequency
discrimination can be employed to achieve this.  However, these requirements can be met for
certain atomic or molecular systems, as we have demonstrated for Rubidium and hydrogen 
fluoride.

\section*{Acknowledgements}
We sincerely thank A.~I.\ Solomon and A.~V.\ Durrant of the Open University for helpful 
discussions and suggestions.  ADG would like to thank the EPSRC for financial support
and VR would like to acknowledge the support of NSF Grant DMS 0072415.  

\appendix 
\section{Derivation of equation (\ref{eq:Omega})}
\label{appendix:A}

Let $\tilde{E}_n=E_n/\hbar$ and $\tilde{d}_n=d_n/\hbar$.
Inserting equations (\ref{eq:U0}) and (\ref{eq:Hm}) into (\ref{eq:SE2}) leads to
\begin{eqnarray*}
\fl \rmi\frac{d\op{U}_I(t)}{dt} 
    &=& \op{U}_0(t)^\dagger\left\{\sum_{m=1}^M\op{H}_m[f_m(t)] /\hbar \right\}
        \op{U}_0(t)\op{U}_I(t)\\
\fl &=& \sum_{n,m,n'} e^{\rmi \tilde{E}_n t} \op{e}_{n,n} 
        A_m(t) \tilde{d}_m \left(e^{\rmi(\omega_m t + \phi_m)}\op{e}_{m,m+1} 
              e^{-\rmi(\omega_m t + \phi_m)}\op{e}_{m+1,m} \right) 
              e^{-\rmi \tilde{E}_{n'} t} \op{e}_{n',n'} \op{U}_I(t)\\
\fl &=& \sum_m A_m(t)\tilde{d}_m \left(e^{\rmi\tilde{E}_mt} e^{\rmi(\omega_m t+\phi_m)} 
         e^{-\rmi\tilde{E}_{m+1} t} \op{e}_{m,m+1}e^{\rmi\tilde{E}_{m+1} t} 
         e^{-\rmi(\omega_m t +\phi_m)}e^{-\rmi\tilde{E}_m t}\op{e}_{m+1,m}\right)
         \op{U}_I(t)\\
\fl &=&\sum_m A_m(t) \tilde{d}_m 
    \left( e^{\rmi\phi_m}\op{e}_{m,m+1} + e^{-\rmi\phi_m}\op{e}_{m+1,m}\right)\op{U}_I(t)\\
\fl &=&\sum_m A_m(t) \tilde{d}_m\left[\cos\phi_m \left(\op{e}_{m,m+1}+\op{e}_{m+1,m} \right)
    +\rmi\sin\phi_m \left(\op{e}_{m,m+1}-\op{e}_{m+1,m} \right) \right] \op{U}_I(t)\\
\fl &=&\sum_m A_m(t) \tilde{d}_m 
    \left( -\rmi\op{y}_m \cos\phi_m + \op{x}_m \rmi\sin\phi_m \right) \op{U}_I(t).
\end{eqnarray*}
Hence, multiplying both sides by $-\rmi$ gives
\begin{equation}
  \frac{d\op{U}_I(t)}{dt} = \sum_m A_m(t) \tilde{d}_m 
  \left(\op{x}_m  \sin\phi_m - \op{y}_m \cos\phi_m \right) \op{U}_I(t). 
\end{equation}

\section{Lie group decomposition algorithm} 
\label{appendix:Udecomp}

To find a decomposition (\ref{eq:Udecomp}) for the unitary operator $\op{U}$ we define 
the equivalent operator $\op{U}^{(0)}\in SU(N)$ by $\op{U}^{(0)}=e^{-\rmi\Gamma/N}\op{U}$
where $e^{\rmi\Gamma}=\det(\op{U})$.  Our goal is to reduce $\op{U}^{(0)}$ step by step to
a diagonal matrix whose diagonal elements are arbitrary phase factors $e^{\rmi\theta_n}$.
Recall that this reduction is always sufficient if the initial state of the system is an 
ensemble of energy eigenstates.  

Let $U_{ij}^{(0)}$ denote the $i$th row and $j$th column entry in the matrix representation
of $\op{U}^{(0)}$.  In the first step of the decomposition we seek a matrix 
\begin{equation}
  \op{W}^{(1)}=\exp\left[-C_1\left(\sin\phi_1\op{x}_1 -\cos\phi_1\op{y}_1\right)\right],
\end{equation}
which is the identity matrix everywhere except for a $2\times 2$ block of the form
\begin{equation}
 \left( \begin{array}{cc} 
       \cos(C_1)                      & \rmi e^{\rmi\phi_1} \sin(C_1) \\
       \rmi e^{-\rmi\phi_1} \sin(C_1) & \cos(C_1)   
  \end{array} \right)
\end{equation}
in the top left corner, such that
\begin{equation} \label{eq:W1}
    \op{W}^{(1)} \left(\begin{array}{c} U_{1,N}^{(0)} \\ 
                                        U_{2,N}^{(0)} \\
                                        \vdots
                  \end{array}\right)
  = \left(\begin{array}{c} 0 \\ c \\ \vdots \end{array}\right)
\end{equation}
where $c$ is some complex number.  Noting that $U_{1,N}^{(0)}=r_1 e^{\rmi\alpha_1}$ and
$U_{2,N}^{(0)} = r_2 e^{\rmi\alpha_2}$, it can easily be verified that setting 
\begin{equation} \label{eq:Cphi}
    C_k    = -\mbox{arccot}(-r_2/r_1), \quad
    \phi_k = \pi/2+\alpha_1-\alpha_2 
\end{equation}
achieves (\ref{eq:W1}).  Next we set $\op{U}^{(1)} = \op{W}^{(1)} \op{U}^{(0)}$ and find
$\op{W}^{(2)}$ of the form
\begin{equation}
  \op{W}^{(2)} = \exp\left[-C_2\left(\sin\phi_2\op{x}_2 -\cos\phi_2\op{y}_2\right)\right]
\end{equation}
such that
\begin{equation} \label{eq:W2}
    \op{W}^{(2)} \left(\begin{array}{c} 0 \\
                                        U_{2,N}^{(1)} \\ 
                                        U_{3,N}^{(1)} \\
                                        \vdots 
                  \end{array}\right)
  = \left(\begin{array}{c} 0 \\ 0 \\ c \\ \vdots \end{array}\right)
\end{equation}
where $c$ is again some complex number.  Repeating this procedure $N-1$ times leads to
a matrix $\op{U}^{(N-1)}$ whose last column is $(0,\ldots,0,e^{\rmi\theta_N})^T$.  Since
we are not concerned about the phase factor $e^{\rmi\theta_N}$ in this paper, we stop 
here.  Note that
\[ 
 \exp\left(-C \op{x}_{N-1}\right) \times 
 \exp\left[-C (\op{x}_{N-1}\sin\phi-\op{y}_{N-1}\cos\phi) \right]
\]
with $C=\pi/2$ and $\phi=-\pi/2-\theta_n$ maps $(0,e^{\rmi\theta_{N-1}})^T$ onto $(0,1)^T$.
Hence, a complete reduction to the identity matrix would require two additional steps to 
eliminate $e^{\rmi\theta_N}$, which would result in two additional control pulses.

Having reduced the last column, we continue with the $(N-1)$st column in the same fashion,
noting that at most $N-2$ steps will be required to reduce the $(N-1)$st column to $(0,
\ldots,0, e^{\rmi\theta_{N-1}},0)^T$ since $\op{U}^{(0)}$ is unitary.   We repeat this 
procedure until after at most $K=N(N-1)/2$ steps $\op{U}^{(0)}$ is reduced to a diagonal
matrix $\mbox{diag}(e^{\rmi\theta_1},\ldots,e^{\rmi\theta_N})$ and we have
\begin{equation}
   \op{W}^{(K)} \cdots \op{W}^{(1)} \op{U}^{(0)} 
 = \mbox{diag}\left(e^{\rmi\theta_1},\ldots,e^{\rmi\theta_N} \right).
\end{equation}
Finally, setting $\op{V}_k \equiv \left(\op{W}^{(K+1-k)}\right)^\dagger$ leads to
\begin{equation}
  \op{U}^{(0)} = \op{V}_K \op{V}_{K-1} \cdots \op{V}_1 
              \mbox{diag}\left(e^{\rmi\theta_1},\ldots,e^{\rmi\theta_N} \right)
\end{equation}
and therefore $\op{U} = \op{V}_K \op{V}_{K-1} \cdots \op{V}_1 \Theta$, where $\Theta=
e^{\rmi\Gamma/N}\mbox{diag}\left(e^{\rmi\theta_1},\ldots,e^{\rmi\theta_N}\right)$ is a 
diagonal matrix of phase factors.

Recall that $\op{U}$ can always be decomposed such that $\Theta$ is the identity matrix.
However, up to $2(N-1)$ additional terms would be required to eliminate the phase factors,
which would result in additional control pulses.  While some applications indeed require 
the elimination of these phase factors, they are often insignificant and the additional 
control pulses would be superfluous.  For a more sophisticated decomposition algorithm 
that requires only very few phases the reader is referred to \cite{CP267p25}.

\section*{References}
\bibliography{papers2000,papers9599,papers8089,papers9094,books,Noordam}
\end{document}